\documentclass[aps,prb,amsmath,superscriptaddress,twocolumn,longbibliography]{revtex4-1}

\usepackage{graphicx}
\usepackage{amsmath}
\usepackage{amsfonts}
\usepackage{amssymb}
\usepackage{braket}
\usepackage{bm}
\usepackage{multirow}
\usepackage{color}
\usepackage[normalem]{ulem}
\usepackage{mathrsfs}
\usepackage{mathtools}
\usepackage{float}

\makeatletter

\newcommand{\Rmnum}[1]{\expandafter\@slowromancap\romannumeral #1@}
\makeatother

\usepackage[breaklinks]{hyperref}
\hypersetup{colorlinks=true, linkcolor=blue, citecolor=blue, filecolor=blue, urlcolor=blue}

\AtBeginDocument{%
    \newwrite\bibnotes
    \def\bibnotesext{Notes.bib}
    \immediate\openout\bibnotes=\jobname\bibnotesext
    \immediate\write\bibnotes{@CONTROL{REVTEX41Control}}
    \immediate\write\bibnotes{@CONTROL{%
    apsrev41Control,author="08",editor="1",pages="1",title="0",year="1"}}
     \if@filesw
     \immediate\write\@auxout{\string\citation{apsrev41Control}}%
    \fi
}%

\def\i{\mathrm i}

\def\sz{\sigma^{\rm z}}
\def\sx{\sigma^{\rm x}}
\def\sy{\sigma^{\rm y}}

\def\sm{\sigma^{-}}
\def\mz{\mu^{\rm z}}
\def\mx{\mu^{\rm x}}
\def\my{\mu^{\rm y}}

\def\mm{\mu^{-}}
\def\tz{\tau^{\rm z}}
\def\tx{\tau^{\rm x}}

\def\tz{\tau^{\rm z}}
\def\tx{\tau^{\rm x}}

\def\hA{\hat a}
\def\hB{\hat b}

\begin{document}

\title{A superconducting circuit realization of combinatorial gauge symmetry}

\author{Claudio Chamon}
\email{chamon@bu.edu}
%\thanks{The authors contributed equally}
\affiliation{Physics Department, Boston University, Boston, MA, 02215, USA}

\author{Dmitry Green}
\email{dmitry.green@aya.yale.edu}
%\thanks{The authors contributed equally}
\affiliation{AppliedTQC.com, ResearchPULSE LLC, New York, NY 10065, USA}

\author{Andrew J.  Kerman}
\email{ajkerman@ll.mit.edu}
\affiliation{Lincoln Laboratory, Massachusetts
  Institute of Technology, Lexington, MA, 02421, USA}

\date{\today}

\begin{abstract}
  We propose a superconducting quantum circuit based on a general
  symmetry principle -- combinatorial gauge symmetry -- designed to
  emulate topologically-ordered quantum liquids and serve as a
  foundation for the construction of topological qubits. The proposed
  circuit exhibits rich features: in the classical limit of large
  capacitances its ground state consists of two superimposed loop
  structures; one is a crystal of small loops containing disordered
  $U(1)$ degrees of freedom, and the other is a gas of loops of all
  sizes associated to $\mathbb{Z}_2$ topological order. We show that
  these classical results carry over to the quantum case, where phase
  fluctuations arise from the presence of finite capacitances,
  yielding ${\mathbb Z}_2$ quantum topological order. A key feature of
  the exact gauge symmetry is that amplitudes connecting different
  ${\mathbb Z}_2$ loop states arise from paths having zero classical
  energy cost. As a result, these amplitudes are controlled by
  dimensional confinement rather than tunneling through energy
  barriers. We argue that this effect may lead to larger energy gaps
  than previous proposals which are limited by such barriers,
  potentially making it more likely for a topological phase to be
  experimentally observable. Finally, we discuss how our
  superconducting circuit realization of combinatorial gauge symmetry
  can be implemented in practice.
  
 \end{abstract}

\maketitle

%%%%%%%%%%%%%%%%%%%%%%%%%%%%%%%%%%%%%%%%%%%%%%%%%%%%%%%%%%%%%%%%%%%%%%%%
%%%%%%%%%%%%%%%%%%%%%%%%%%%%%%%%%%%%%%%%%%%%%%%%%%%%%%%%%%%%%%%%%%%%%%%%
\section{Introduction}
\label{sec:intro}

%%%%%%%%%%%%%%%%%%%%%%%%%%%%%%%%%%%%%%%%%%%%%%
\begin{figure*}[!th]
\centering
\includegraphics[width=1\textwidth]{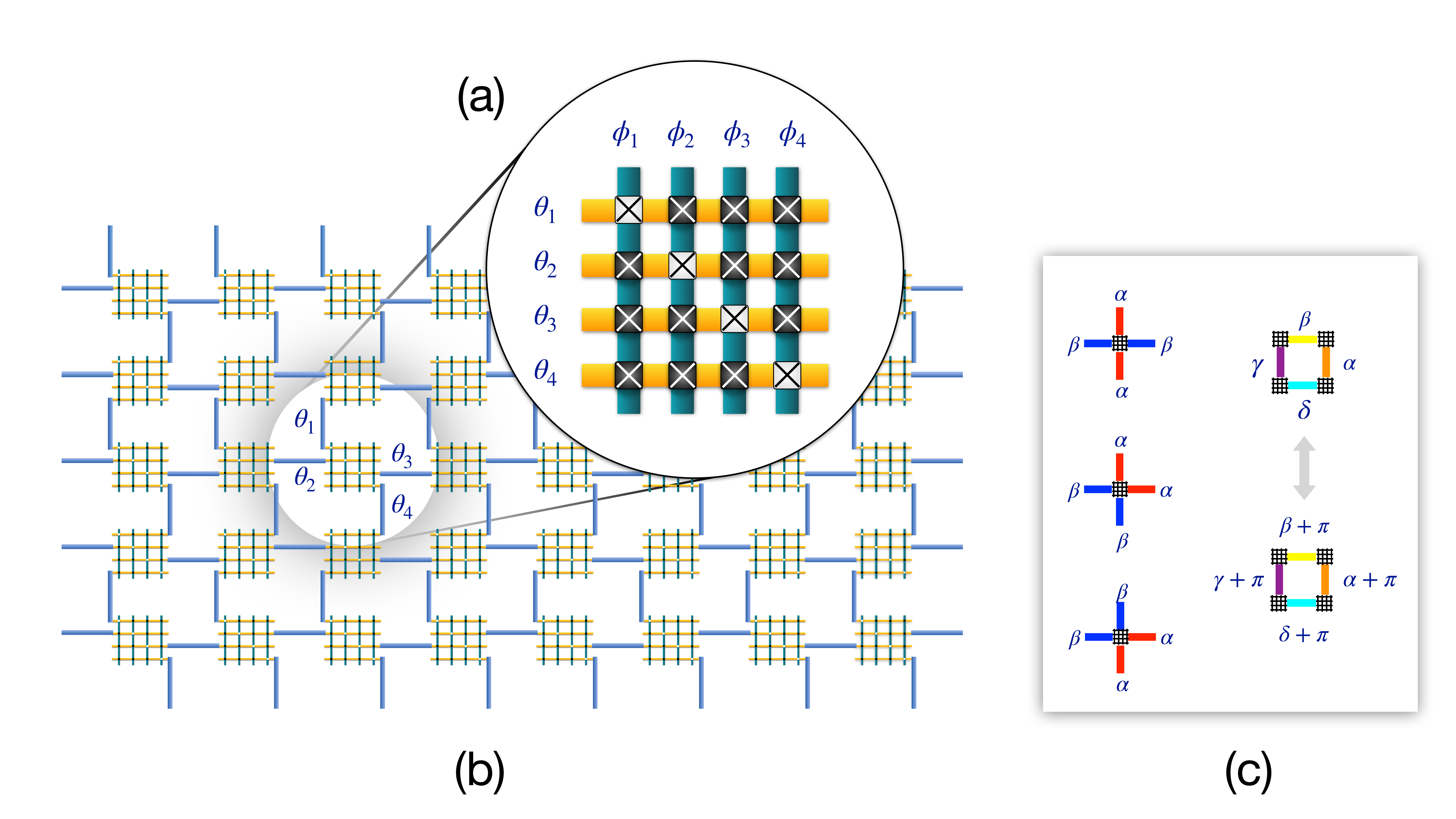}
\caption{(a) An array of intersecting superconducting wires which form
  a single ``waffle" site in the lattice. Vertical (green) wires are
  ``matter" degrees of freedom with phase $\phi_n$ and charge
  $q_n$. Horizontal (yellow) wires are ``gauge" degrees of freedom
  with phase $\theta_i$ and charge $Q_i$. Wires are coupled by
  Josephson junctions depicted as black or white crosses depending on
  the sign of the coupling. This array mirrors the interaction matrix
  $W$ and implements combinatorial gauge symmetry. (b) The full
  lattice with links (blue) connecting the gauge degrees of
  freedom. One site is highlighted to illustrate how sites are
  connected; links that it shares with neighboring sites are labeled
  by their gauge phases $\theta_i$. The matter wires with phases
  $\phi_n$ are connected only to gauge wires (c) This geometry leads
  to \textit{exact} $\mathbb{Z}_2$ topological order, as illustrated
  by the allowed configurations of the gauge phases $\theta$ emanating
  from each waffle site. We show the limit where the capacitances
    are large (classical) limit where the gauge phases are good
    quantum numbers. Gauge phases must be equal pairwise $\alpha$
    (red) and $\beta$ (blue) at each site, where both phases are
    defined modulo $\pi$. In the context of lattice models, such
    vertices are related to loop models. Additionally, plaquettes may
    be flipped back and forth by shifting all gauge phases around the
    plaquette $(\alpha, \beta, \gamma, \delta)$ by $\pi$.}
\label{fig:JJ_lattice} 
\end{figure*}
%%%%%%%%%%%%%%%%%%%%%%%%%%%%%%%%%%%%%%%%%%%%%%

Quantum circuits based on Josephson
junctions~\cite{devoret2004superconducting} have increasingly
leveraged the techniques of large-scale integrated circuit fabrication
in recent years, and this technology has become the basis for the
largest quantum information processing systems demonstrated to-date
\cite{DWaveBKT,Arute_2019,stateofplay}. These circuits can also be engineered
to emulate physical quantum systems and basic phenomena, such as the
Berezinskii-Kosterlitz-Thouless transition in the
XY-model~\cite{Resnick-etal}. The goal of this paper is to describe a
superconducting quantum circuit based on a symmetry principle --
combinatorial gauge symmetry~\cite{CGS} -- which can be used to
realize topologically ordered states in an engineered quantum system.

The study of topologically ordered states of matter~\cite{Wen1990a}
remains an active area of research in condensed matter physics. This
class of states includes, for instance, quantum spin
liquids~\cite{Savary2016}, which are devoid of magnetic
symmetry-breaking order but display topological ground state
degeneracies. A number of solvable spin models exist as examples, but
these theoretical models include multi-spin interactions not realized
in nature. One notable exception of a model with only two-body
interactions is the Heisenberg-Kitaev
model~\cite{hex_Kitaev,Jackeli2009}, but its realization in a material
system appears to reside within its non-topological phase.

As opposed to seeking naturally occurring materials, here we follow a
similar route to that of Refs.~\cite{Ioffe1, Ioffe2, Doucot2005,
  Ioffe2003, Gladchenko, Doucot2012}, and focus on engineering
topologically ordered systems using superconducting quantum
circuits. In the models considered in those works, a gauge symmetry
emerges in the limit where the Josephson energy is dominant and the
superconducting phase is the good quantum number. Once the correct
manifold of states is selected through the Josephson coupling, quantum
phase fluctuations induced by the charging energy give rise to a
perturbative energy gap that stabilizes the topological phase. The
main issue with this emergent symmetry is that it only holds in the
perturbative regime where the Josephson energy is much larger than the
charging energy.

While the emergent symmetry ensures the existence of the topological
phase, its intrinsically perturbative nature fundamentally limits the
size of the gaps that can be obtained. One possible way to escape
these limits is to design a system for which the gauge symmetry is
{\it exact at the microscopic level} and therefore non-perturbative,
holding for {\it any} strength of the coupling constants, including
regimes where the charging energy dominates. Such an exact symmetry
should therefore expand the range of parameters for which the
topological phase may be stable. In this paper we present a proposal
for such a system, in the form of a quantum circuit that exhibits
exact combinatorial gauge symmetry, including a proposal for how to
realize this circuit experimentally.

From a purely theoretical perspective, combinatorial gauge symmetry is
interesting in its own right. It can be applied to spins, fermions, or
bosons, and all these systems show rich behaviors as a result of the
symmetry. We shall present examples of superconducting XY-like systems
with coexisting $U(1)$ and ${\mathbb Z}_2$ loop structures. These two
loop structures arise from the form of the designed Josephson
couplings: the superconducting phases are locked around the $U(1)$
loops, but only mod $\pi$ (not $2\pi$), as these phases can be shifted
by $\pi$ along the closed paths of the ${\mathbb Z}_2$ loops without
changing either the Josephson or the electrostatic energy. We show
that the $U(1)$ structure crystallizes into an array of small loops
while the ${\mathbb Z}_2$ structure forms a gas of loops at all
scales. These XY-like systems, unlike the usual XY-model, do not show
quasi-long-range order of the $U(1)$ degrees of freedom, precisely
because of the local loop structures. However, the ${\mathbb Z}_2$
degrees of freedom realize a topologically ordered state in the same
class as in the toric or surface codes, and hence the quantum circuits
presented here can be used for building topological qubits.

The paper is organized as follows. In Sec.~\ref{sec:array} we
introduce the superconducting circuit that realizes combinatorial
gauge symmetry, and we summarize the key elements of this symmetry.
In Sec.~\ref{sec:JJ-general} we show how topological features
naturally arise in the classical limit of large capacitances, in the
form of both $U(1)$ and $\mathbb Z_2$ loop structures. In
Sec.~\ref{sec:JJ-semi} we discuss how quantum fluctuations endow the
loops with dynamics, and how the quantum system is described by an
effective toric/surface code Hamiltonian. Finally, in
Sec.~\ref{sec:realization} we present a detailed discussion of
realistic circuit elements needed for an experimental construction.

\section{Superconducting wire array with combinatorial gauge symmetry}
\label{sec:array}
The array of superconducting wires we consider are depicted in
Fig.~\ref{fig:JJ_lattice}. Looking at a given site in
Fig.~\ref{fig:JJ_lattice}(a), each of the four vertical wires is
coupled to each of the four horizontal wires by a Josephson junction
in a kind of ``waffle" geometry. The waffles are placed at the sites
of a square lattice, as shown in Fig.~\ref{fig:JJ_lattice}(b), and are
labeled by $s$. The ``matter" wires with superconducting phase
$\phi_n$ are confined to each waffle (or site), and they are indexed
by $n$, with $n\in s$ denoting the set of four wires in waffle
$s$. The ``gauge" wires with phase $\theta_i$ are shared between
sites, spanning the links or bonds of the square lattice, labeled by
$i$, with $i\in s$ denoting the set of four links emanating from site
$s$. Each of these phases has a conjugate dimensionless charge
variable, satisfying the commutation relations
$ [\phi_n,q_{m}] = \i\,\delta_{mn}$ and
$[\theta_i,Q_{j}] = \i\,\delta_{ij}$.

The Hamiltonian for the system is composed of electrostatic (kinetic) and Josephson (potential) terms:

\begin{align}
  H = H_K + H_J 
  \;.\label{eq:Htot}
\end{align}

\noindent The kinetic energy is given by:

\begin{align}
  H_{K}&=\frac{1}{2}\vec{\mathbf Q}^{\rm T}\cdot {\mathbf C}^{-1}\cdot\vec{\mathbf Q}
  \label{eq:HK}
\end{align}

\noindent where ${\mathbf C}^{-1}$ is the system's inverse capacitance
matrix and $\vec{\mathbf Q}$ is a vector containing all of the island
charges, so that if we define
\begin{align}
\vec{\mathbf Q}&\equiv 2e\begin{pmatrix}
      \vec{Q} \\
      \vec{q} \\
\end{pmatrix},\;
\vec{\mathbf \Phi}\equiv\frac{\Phi_0}{2\pi}\begin{pmatrix}
      \vec{\theta} \\
      \vec{\phi} \\
\end{pmatrix}\label{eq:Qphi}
\end{align}
with $e$ the electron charge and $\Phi_0\equiv h/2e$ the
superconducting fluxoid quantum, the canonical commutation relations
can be re-written as
$[\vec{\mathbf\Phi},\vec{\mathbf Q}]=\hbar \openone$.

The Josephson potential is given by
%\begin{subequations}
\begin{subequations}
\label{eq:JJ-J}
\begin{align}
&
  H_J = -J\;\sum_s \left[\sum_{n,i\in s} W_{ni}\;\cos(\phi_n-\theta_i)\right]
\end{align}
where we take $J>0$. The core component is the $4\times 4$ interaction
matrix $W$, which is what enables the combinatorial symmetry and
drives the physical connectivity of the circuit. It is required to be
a so-called Hadamard matrix whose elements are $\pm1$ and it is
orthogonal $W^\top W=4 \openone$. A convenient choice is
\begin{align}
  W =
  \begin{pmatrix}
    -1&+1&+1&+1\\
    +1&-1&+1&+1\\
    +1&+1&-1&+1\\
    +1&+1&+1&-1
  \end{pmatrix}
  \;,
  \label{eq:W}
\end{align}
\end{subequations}
and all other choices are physically equivalent. The coupling matrix is captured literally by the waffle geometry in Fig.~\ref{fig:JJ_lattice}(a).  Hadamard
matrices are invariant under a group of monomial transformations,
which is the source of the gauge symmetry. Specifically, we have the
automorphism
\begin{align}
  L^{-1}\;W\;R = W
  \;,
  \label{eq:automorphism}
\end{align}
where $R$ and $L$ are monomial matrices -- generalized permutation
matrices with matrix elements $\pm 1$ or 0. Monomial transformations
preserve the commutation relations of the underlying
operators~\cite{CGS}, which in this case are the phases and charges on
all wires.  For example, with our choice of $W$ in Eq.~(\ref{eq:W}),
the following pair satisfies the automorphism~(\ref{eq:automorphism})
on \textit{each site} $s$:
\begin{align}
L=&
\begin{pmatrix}
0 & +1 & 0 & 0 \\
+1 & 0 & 0 & 0 \\
0 & 0 & 0 & -1 \\
0 & 0 & -1 & 0 
\end{pmatrix}
&
R=&
\begin{pmatrix}
-1 & 0 & 0 & 0 \\
0 & -1 & 0 & 0 \\
0 & 0 & +1 & 0 \\
0 & 0 & 0 & +1 
\end{pmatrix}
\label{eq:LR}
\;.
\end{align}
Here it looks like we are only transforming the interaction, but of
course quantum mechanically transforming an operator is equivalent to
transforming the state. In this case the $R$ matrix acts on the phases
of the gauge wires $\theta_i$ on a given site, shifting the phase by
$\pi$ whenever there is a $-1$. Similarly, $L$ acts on the phases of
the matter wires $\phi_n$, shifting them by $\pi$ whenever there is a
$-1$, and in such a way as to preserve the required symmetry
(\ref{eq:automorphism}) on each site. The key is that the matter wires
are only connected locally on each site, hence their phases may be
permuted as well as shifted in general. The gauge wires, on the other
hand, bridge two waffles, and therefore the gauge phases can be
shifted but not permuted, and hence the matrix $R$ must be diagonal.

%In other words, matter wires serve to entangle the states at each site.

%%%%%%%%%%%%%%%%%%%%%%%%%%%%%%%%%%%%%%%%%%%%%%
\begin{figure}[!tbh]
\centering
\advance\leftskip-0.25cm
\includegraphics[width=0.45\textwidth]{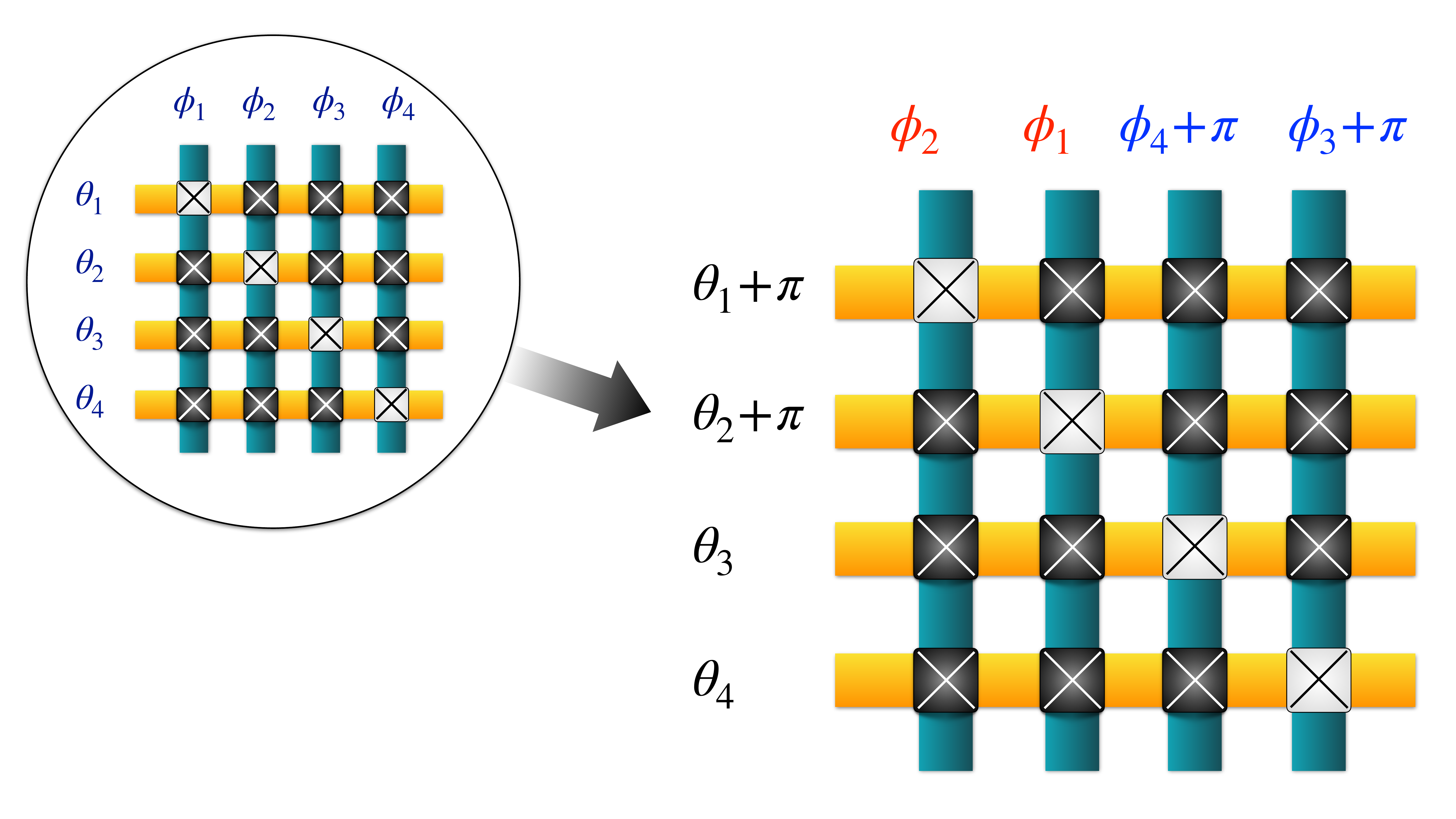}
\caption{Visual illustration of the action of an $L$ and $R$ pair on
  the many-body states at each site. The pictured transformation
  corresponds to the automorphism in Eq.~(\ref{eq:LR}) applied to a
  generic state of the waffle (circle). Note that the signs of the
  junctions replicate the pattern of $\pm 1$'s in $W$. Wires and
  couplings are fixed; only the phases are being permuted and/or
  shifted. Combinatorial gauge symmetry guarantees that the
  Hamiltonian is invariant under this transformation. The phases of
  the matter wires can be shifted by $\pi$ \textit{and} permuted as
  they only live on that site (permutations indicated by red and blue
  $\phi$'s). However, states of the gauge wires can be shifted but
  \textit{not} permuted as they are shared by neighboring sites. Hence
  $R$ must be diagonal, but $L$ need only be a monomial matrix. When
  connected on the lattice, shifting a gauge phase on one site
  automatically shifts it on its neighboring site, resulting in
  degenerate states that are loops on the lattice.}
\label{fig:CGSexample} 
\end{figure}
%%%%%%%%%%%%%%%%%%%%%%%%%%%%%%%%%%%%%%%%%%%%%%

The fact that the extra permutation symmetry is \textit{local} is
crucial and gives rise to the topological nature of the waffle
circuit. The topological structures that arise in this circuit are
illustrated in Fig.~\ref{fig:JJ_lattice}(c) and discussed in detail
in Secs.~\ref{sec:JJ-general} and \ref{sec:JJ-semi} below.

Thus far we concentrated on the Josephson couplings in the potential
energy term; the capacitance matrix can be quite general for the
properties we discuss in the paper, provided the it is symmetric under
the permutation of the matter wires within a waffle $s$. Basically,
this requirement ensures that $H_K$ is invariant under the permutation
part of the transformation associated with the $L$ matrices such as
those in Eq.~\eqref{eq:LR}. We present an experimental setting for
such symmetry condition to hold in Sec.~\ref{sec:realization}.

%\subsection{Combinatorial Gauge Symmetry - generalized framework}

Before proceeding with an analysis of the waffle superconducting array, we summarize the mathematical foundation for why it realizes combinatorial gauge symmetry. The general structure will simplify our analysis. And, we will see that the waffle array is a special case, so that the approach can be used to construct other kinds of systems with combinatorial gauge symmetry.

In the most general case, we can write an interaction of the form
\begin{align}
  H_J = -J\;\sum_s \sum_{n,i\in s} W_{ni}\;
  \left(
  \hA^\dagger_n\;\hB^{}_i
  +
  \hB^\dagger_i \;\hA^{}_n
  \right)
  \;,
  \label{eq:JJ-general}
\end{align}
where $\hA^{}_n$ and $\hB^{}_i$ are generic degrees of freedom.  In
fact we can use any angular momentum, fermionic, or bosonic
variables. (Noticeably, when used as a hopping amplitudes for bosons
or fermions, the $W$ matrix yields flat bands.) An essential feature
is that the $\hA_n$ are ``matter" fields localized to each site which
enables us to use permutation symmetry without distorting the
lattice. The $\hB_i$ are ``gauge" fields which are shared by lattice
sites $s$.

According to the automorphism symmetry of $W$ that we have already
introduced in Eq.~(\ref{eq:automorphism}) the operators $\hA$ and
$\hB$ transform as
\begin{align}
  \hA_n \to \sum_m \hA_m\;(L^{-1})_{mn}
  \quad
  \text{and}
  \quad
  \hB_i \to \sum_j R_{ij}\;\hB_j
  \;.
  \label{eq:transform_LR}
\end{align}
To implement the sign changes in the monomial symmetries, such as
those in Eq.~(\ref{eq:LR}) we require that there exist unitary
transformations $U^{(L)}_n$ and $U^{(R)}_i$ such that
\begin{align}
  U^{(L)}_n\,\hA_n^{\phantom\dagger}\,U^{(L)\dagger}_n = -\hA_n^{ }
  \;\;
  \text{and}
  \;\;
  U^{(R)}_i\,\hB_i^{\phantom\dagger}\,U^{(R)\dagger}_i = -\hB_i^{ }
  \;.
  \label{eq:U}
\end{align}
These sign-flip transformations, when combined with permutations of
the $n$ and $i$ indices, lead to the monomial transformations written
in Eq.~(\ref{eq:transform_LR}), which preserve the proper commutation
relations of the $\hA$ and $\hB$ operators. We refer to
Ref.~\onlinecite{CGS} for the special case of how to realize the
$\mathbb{Z}_2$ gauge theory or toric code using spin-1/2.

To this Hamiltonian $H_J$ one can add any kinetic term $H_K$ that
commutes with the unitary operators $U^{(L)}_n$ and $U^{(R)}_i$, and
that have couplings that are independent of $n$ and $i$, so that
  permutation invariance holds. In the particular case that the $R$
  transformation matrices are restricted to be diagonal, then the
  couplings need only be independent of $n$, so that the permutation
  part of the $L$ transformations, see Eq.~\eqref{eq:LR}, leaves $H_K$
  altogether invariant. When these conditions are satisfied, the
whole Hamiltonian obeys combinatorial gauge symmetry.

The superconducting wire array is an example of this general
framework. In the Hamiltonian with kinetic and potential terms in
Eqs.~(\ref{eq:HK}) and (\ref{eq:JJ-J}) we identify the matter and
gauge fields as the phases of the superconducting wires, as follows:
\begin{align}
  \hA_n = e^{\i\, \phi_n}
  \quad
  &\text{and}
  \quad
  \hB_i = e^{\i\, \theta_i}
  \;,
  \nonumber\\
  U^{(L)}_n = e^{\i\pi\, q_n}
  \quad
  &\text{and}
  \quad
  U^{(R)}_i = e^{\i\pi\, Q_i}
  \;.
  \label{eq:JJ_full}
\end{align}
$U^{(L)}_n$ and $U^{(R)}_i$ are generated by the conjugate variables
$q_n$ and $Q_i$, respectively, hence they commute with the kinetic
term. The action of $U^{(L)}_n$ and $U^{(R)}_i$ on $\hA_n$ and $\hB_i$
can be thought of as shifting $\phi_n$ or $\theta_i$ by $\pi$. So in
addition to the usual global symmetry that shifts all phases equally,
we have a \textit{local} symmetry that shifts an even number of
$\theta_i$'s and $\phi_n$'s by $\pi$ in each {\it star}, i.e., each
site $s$ with its four links $i\in s$. This transformation can be done
consistently on four neighboring stars at the corners of a {\it
  plaquette} $p$; the resulting transformation shifts the phase of the
four links on the edges (gauge wires) of the plaquette by $\pi$, along
with the corresponding transformations of the matter wires. This
transformation is associated with a local ${\mathbb Z}_2$ symmetry,
and we illustrate this operation in Fig.~\ref{fig:JJ_lattice}(c), on
the right.

%%%%%%%%%%%%%%%%%%%%%%%%%%%%%%%%%%%%%%%%%%%%%%%%%%%%%%%%%%%%%%%%%%%%%%%%

\section{Classical loop model}
\label{sec:JJ-general}

%%%%%%%%%%%%%%%%%%%%%%%%%%%%%%%%%%%%%%%%%%%%%%
\begin{figure*}[!tbh]
\centering
\includegraphics[width=0.75\textwidth]{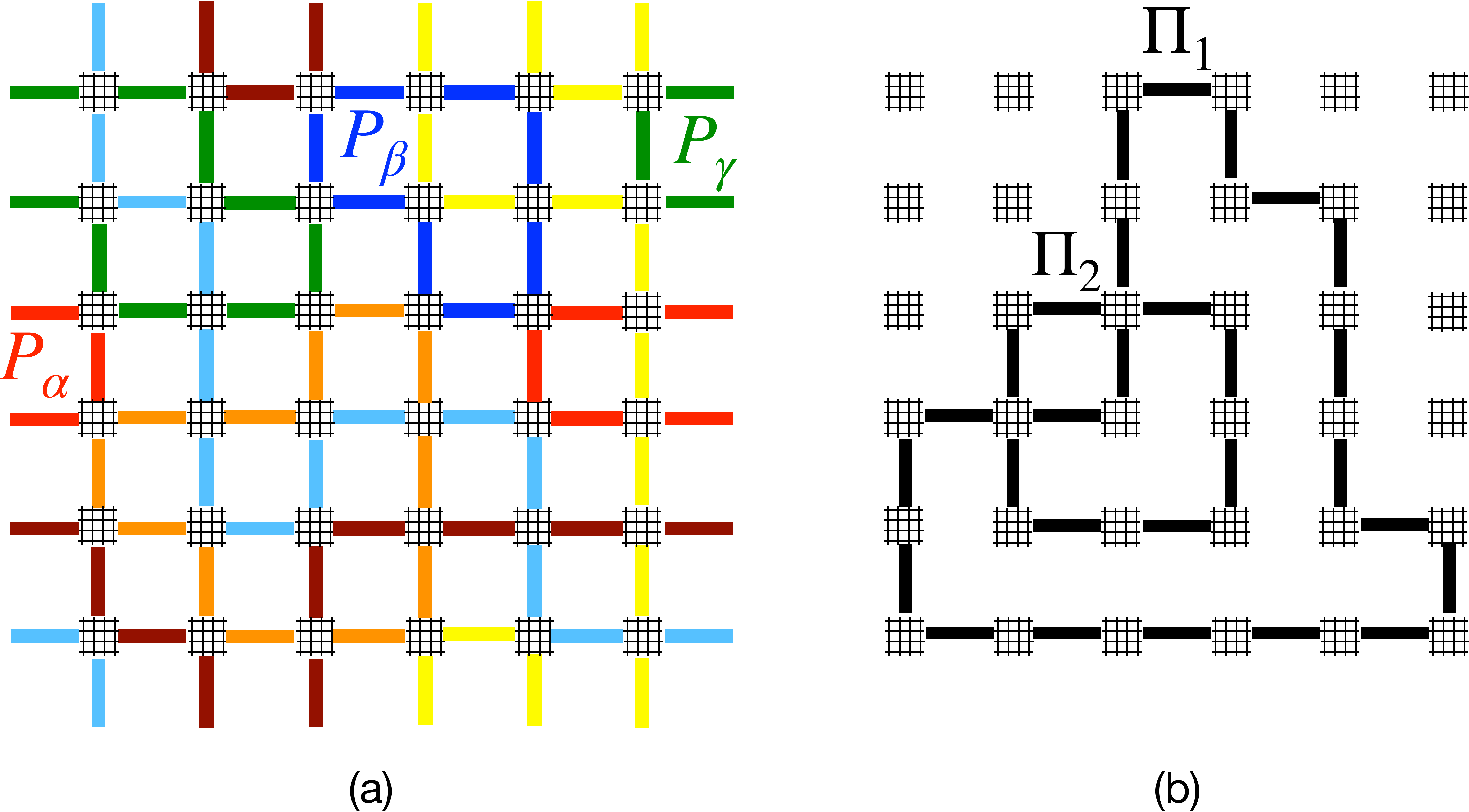}
\caption{(a) Example of the $U(1)$ loop covering in the
  superconducting wire array.  Values of equal $\theta$ are
  represented by different colors, and loops wrap around with periodic
  boundary conditions (e.g., $\theta=\alpha$, $\beta$ and $\gamma$
  label three loops $P$ colored red, blue, and green,
  respectively). Other colored loops fully fill the lattice. Loops can
  intersect. (b) Example of the second kind of loop $\Pi$, which
  adds a phase $\pi$ to each of the links visited. Any such loop is
  a direct result of the local $\mathbb{Z}_2$ gauge symmetry and can
  be superimposed on {\it any} loop covering in (a).}
\label{fig:JJ_paths} 
\end{figure*}
%%%%%%%%%%%%%%%%%%%%%%%%%%%%%%%%%%%%%%%%%%%%%%

We shall show below that a model of loops is realized by the
superconducting wire array with combinatorial gauge symmetry. At the
minima of $H_J$ in Eq.~(\ref{eq:JJ-J}), the $\phi_n$'s in a waffle $s$
become tethered to the $\theta_i$'s:
\begin{equation}
e^{i\phi_n} = \frac{\sum_{i\in s} W_{in} \;e^{i\theta_i}}{\left|\sum_{i\in s}
  W_{in} \;e^{i\theta_i}\right|},
\label{eq:phi}
\end{equation}
with the minimum energy given by
\begin{equation}
E_{\rm min}
  =
-J \sum_s \sum_{n\in s} \left|\sum_{i\in s} W_{ia}
\;e^{i\theta_i} \right|
\;.
\label{eq:minE}
\end{equation}
The manifold of minima is such that $\theta_i$'s and $\phi_n$'s are
equal pairwise at each star. On a given site $s$ let us use the
short-hand ${\bm\theta}=(\theta_1,\theta_2,\theta_3,\theta_4)$ and
similarly for ${\bm\phi}$. Then, for instance, the following minima
have ground state energy $-8J$ at each site:
\begin{subequations}
\begin{align}
{\bm\phi}=(\beta, \alpha, \beta, \alpha)
\quad
  &\text{and}
\quad
{\bm\theta}=(\alpha, \beta, \alpha, \beta),
\label{eq:min1}
\end{align}
where $\alpha$ and $\beta$ are any two phases between $0$ and
$2\pi$. Moreover, we still have the $\mathbb{Z}_2$
symmetry. For example, applying the symmetry operation in
(\ref{eq:LR}) to (\ref{eq:min1}) produces another type of minimum,
\begin{align}
{\bm\phi}=(\alpha, \beta, \alpha+\pi, \beta+\pi)
\quad
  &\text{and}
\quad
{\bm\theta}=(\alpha+\pi, \beta+\pi, \alpha,
\beta).
\label{eq:min2}
\end{align}
\end{subequations}
There are additional minima obtained by symmetry, and their complete
set on each site can be visualized as shown in
Fig.~\ref{fig:JJ_lattice}(c). On the entire lattice, these minima must
be consistent so the ground states are described by loops as we depict
in Fig.~\ref{fig:JJ_paths} with different colors. The lattice
Hamiltonian is confined to the valley of minima as long as the phases
on the four legs at each site are equal pairwise. Therefore, any
fully-packed loop covering -- where each site is visited by two
loops and each loop has the same phase along its path -- will minimize
$H_J$ of Eq.~(\ref{eq:JJ-J}). This class of lattice covering is
associated to $U(1)$ or continuous phases, as depicted in
Fig.~\ref{fig:JJ_paths}(a). In addition, there is another class of
loops, associated to the $\mathbb{Z}_2$ gauge symmetry. The latter
loops do not need to cover all links, but in those links that they do
visit, they shift the phases by $\pi$:
\begin{align}
  \theta_i\to \theta_i + \frac{\pi}{2}(1-\tz_i),
  \;\;
  \tz_i=\pm 1,
  \label{eq:shifts}
\end{align}
where $\tz_i=-1$ indicates that a $\pi$ phase shift is added to link
$i$, while $\tz_i=+1$ indicates no phase shift to that link. These
values can be thought of as the eigenvalues $\pm 1$ of $\tz_i$
operators. Fig.~\ref{fig:JJ_paths}(b) depicts the loops of this second
kind, or $\mathbb Z_2$ loops, which follow the sequence of links $i$
with $\tz_i=-1$. We remark that these loops can be generated starting
from a reference configuration by the application of generators of the
local combinatorial gauge symmetry, plaquette operators [see
Eq.~(\ref{eq:plaquette}) for the general case]:
\begin{align}
  G_p=\prod_{i\in p} \,e^{\i\pi\, Q_i}\,=\,e^{\i\pi\left(\,\sum_{i\in p}Q_i\right)}
  \,=\,\prod_{i\in p}\,\tx_i
  \;.
    \label{eq:plaquette_small}
\end{align}
where the operators $\tx_i$ flip between the eigenvalues $\pm 1$ of
$\tz_i$ operators. Because of the local $\mathbb{Z}_2$ symmetry, the
$U(1)$ phases of the first kind of loops can be seen as defined mod
$\pi$ (rather than $2\pi$), as illustrated by Eqs.~(\ref{eq:min1})
and~(\ref{eq:min2}). Formally, we are working with elements of
$U(1)/{\mathbb Z}_2\cong U(1)$.

In the classical limit of infinite capacitances, we can study the
statistical mechanics of the two loop models where the only energy is
the $H_J$ term of Eq.~(\ref{eq:JJ-J}). Even the $T=0$ limit of the
model is interesting, in that there is a ground state entropy because
of the different ways to cover the lattice with the $U(1)$ and
${\mathbb Z}_2$ loops. Because these two kinds of loops are
independent, the partition function factorizes into the partition
functions of two loop models:
\begin{align}
  Z(T=0)=Z^{U(1)}_{\rm loop}\times Z^{{\mathbb Z}_2}_{\rm loop}
  \;.
\end{align}

The second component corresponds to the usual $\mathbb{Z}_2$ gauge theory. The $U(1)$ component turns out to belong to a class of statistical mechanics models that have been studied in other contexts, such as polymers and lattice spins~\cite{Chayes, Read, Blote, Nahum}. Our case corresponds to the so-called $\text O(N)$ loop model, where $N$ is the number of allowed flavors or colors of each loop. Since we have an infinite set of colors our case is the limit $N\to\infty$.

The zero-temperature partition function accounts for all the
states that minimize the energy, and encodes the entropic contribution
of all allowed loop coverings, hence we can write:
\begin{align}
  Z^{U(1)}_{\rm loop}
  =
  \sum_{\rm loop\;coverings}\;\lambda^{n_\ell}
  \;,
\end{align}
where $\lambda$ is the loop fugacity and $n_\ell$ is the number of
loops in a given loop covering. Since each loop covering is fully
packed, the energy associated to loop length is the same for each
covering, so we have left an overall ground state energy factor out of
the partition function.

We claim that $\lambda\to \infty$ at zero temperature.  Intuitively,
this is because each closed loop can have an infinite number of colors
(continuous phases), so $\lambda$ can be identified with $N$ in this
limit. The intuition is made precise by the following counting
argument. Take a closed loop visiting $p$ sites and $p$ links. The
condition that the phases $\alpha_p$ at each site are equal pairwise
can be viewed as a series of $p$ Boltzmann weights at some divergent
energy scale.  However, only $p-1$ constraints are needed because if
$\alpha_1=\alpha_2=\dots=\alpha_p$ then automatically
$\alpha_p=\alpha_1$ for a loop. In the limit where the Boltzmann
weights become delta functions, the redundant constraint diverges at
zero temperature (formally it is an extra delta function
``$\delta(0)$"). In the Appendix~\ref{sec:appendix_fugacity} we give
a simple example to clarify this argument.

%The $\text O(N)$ model in the limit $N\to\infty$ was also discussed in a more general context~\cite{Chayes, Blote}.

\begin{figure}[!tbh]
\centering
\advance\leftskip-0.25cm
\includegraphics[width=0.5\textwidth]{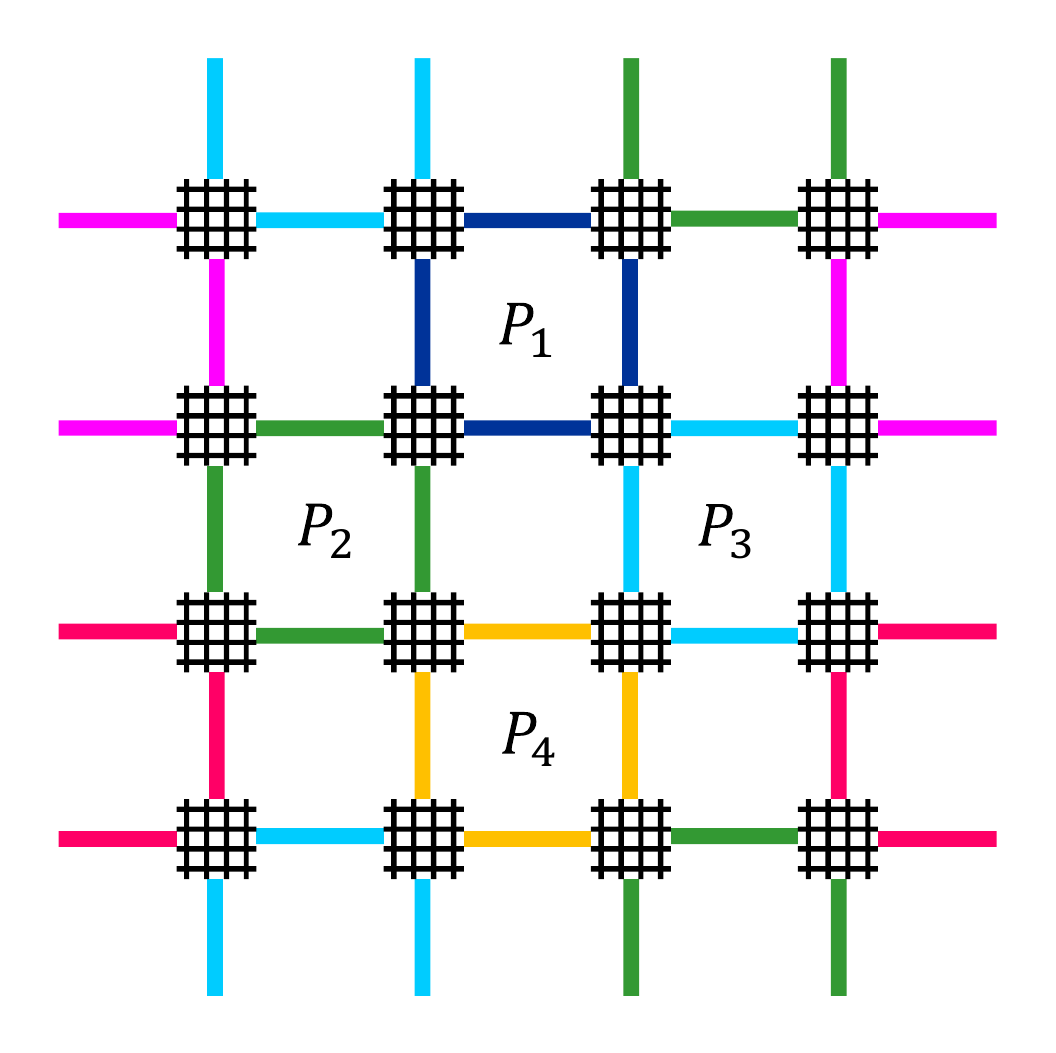}
\caption{Because the phases defined on the loops vary continuously and
  independently, the largest entropy configuration maximizes their
  number; hence it maximizes the number of loops or, equivalently,
  minimizes the length of each loop. This configuration can be thought
  of as independently fluctuating phases around loops on alternating
  plaquettes (depicted by different colors). That this loop crystal
  has maximum entropy originates from the fact that the ``colors'' or
  phases are continuous variables.}
\label{fig:entropy} 
\end{figure}

Due to the infinite fugacity, the system is driven by entropy to
maximize the number of $U(1)$ loops, which is the configuration
illustrated in Fig.~\ref{fig:entropy}. This ground state is the set of
degenerate loop coverings each of which consists of elementary loops
of arbitrary phase around every other plaquette.  There is no
long-range (or quasi-long-range) order of the $U(1)$ loops even at
zero temperature. Since the ground state is dominated by small loops,
any two links further than one lattice spacing belong to distinct
loops and their phases are uncorrelated.

The $\mathbb{Z}_2$ loops on the other hand form a loop gas just like
in the classical limit of the toric
code~\cite{Castelnovo-classical-TO}. Because the long loops in the
$U(1)$ component are exponentially suppressed, they do not destroy the
gapped topological order of the $\mathbb{Z}_2$ component.

%%%%%%%%%%%%%%%%%%%%%%%%%%%%%%%%%%%%%%%%%%%%%%
%%%%%%%%%%%%%%%%%%%%%%%%%%%%%%%%%%%%%%%%%%%%%%

%%%%%%%%%%%%%%%%%%%%%%%%%%%%%%%%%%%%%%%%%%%%%%
\section{Quantum loop model: \\ Emulation of the toric code}
\label{sec:JJ-semi}

The loop models in the previous section originated from the
constraints posed on the superconducting phases at the minimum of the
Josephson energy for the couplings given by the Hadamard matrix
$W$. To endow these loop structures with dynamics, we move away from
the classical limit of infinite capacitances. The finite capacitances
introduce quantum fluctuations to the superconducting phases $\phi_n$
and $\theta_i$ via the kinetic energy expressed in terms of the
conjugate variables $q_n$ and $Q_i$. We shall derive an effective
quantum Hamiltonian describing the dynamics of the $\mathbb{Z}_2$
loops in terms of the $\tz_i$ and $\tx_i$ degrees of freedom discussed
in Eqs.~\eqref{eq:shifts} and \eqref{eq:plaquette_small}.

%%%%%%%%%%%%%%%%%%%%%%%%%%%%%%%%%%%%%%%%%%%%%%
\begin{figure}[!tbh]
  \centering \advance\leftskip-0.25cm
  \includegraphics[width=0.5\textwidth]{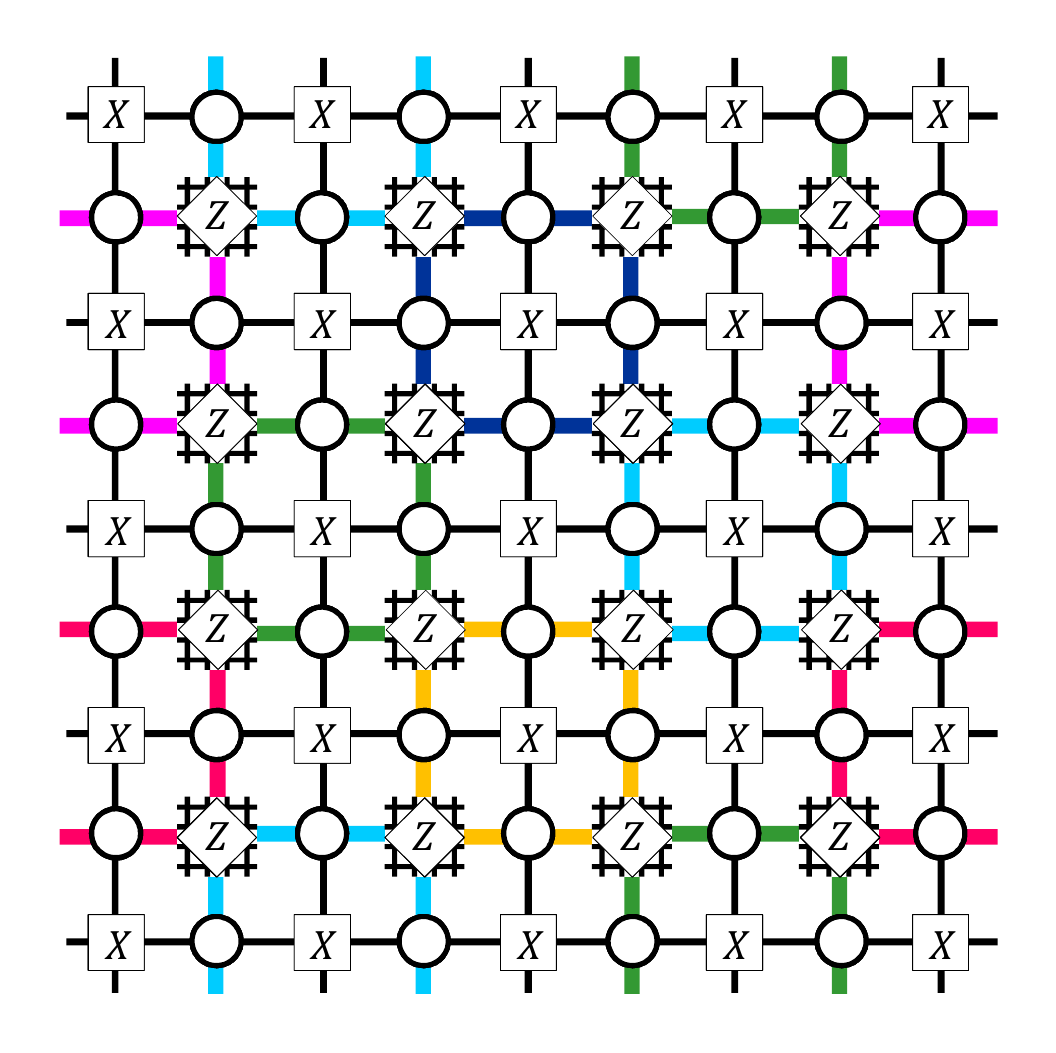}
\caption{The same lattice as Fig. ~\ref{fig:entropy}, but with the mapping to the toric code superposed. The qubits in the toric code are indicated by open circles, and are on the links of the lattice. The mapping between the $U(1)$ loops and the toric code star terms is indicated by the label $Z$, and the mapping between the $\mathbb{Z}_2$ loops and the toric code plaquette terms is indicated by the label $X$.}
\label{fig:toric} 
\end{figure}
%%%%%%%%%%%%%%%%%%%%%%%%%%%%%%%%%%%%%%%%%%%%%%

The link variables $\tz_i$ play the role of quantum spins, whose two
states correspond to the presence or absence of an additional $\pi$
shift on a link. The Josephson coupling penalizes an odd number of
$\pi$ shifts on a star, i.e., configurations with $\prod_{i\in s}
\tz_i=-1$, which is captured in the effective term
\begin{align}
  H^\text{eff}_{\rm star} =  -\lambda_J \sum_s \prod_{i\in s} \tz_i
  \;,
  \label{eq:star-term}
\end{align}
where $\lambda_J=8J$ is the energy separation between the minimum of
Eq.~(\ref{eq:minE}) (which satisfies $\prod_{i\in s} \tz_i=+1$) and a
configuration with an odd number of $\theta_i$ variables shifted by
$\pi$. $H^\text{eff}_{\rm star}$ is the equivalent of the star term in
the toric code.

The effective term governing the flip of an elementary (smallest) loop
plaquette $p$ is written as 
\begin{align}
  H^\text{eff}_{\rm plaquette} = -\sum_p \lambda_{\rm flip}(p) \prod_{i\in
    p} \tx_i \;,
  \label{eq:plaquette-term}
\end{align}
which corresponds to the flip operation on a plaquette depicted in
Fig.~\ref{fig:JJ_lattice}(c), on the right. In
Eq.~\eqref{eq:plaquette-term} we allowed the flipping amplitude
$\lambda_{\rm flip}(p)$ to depend on the plaquette $p$, as we detail
below, along with how to obtain the scale $\lambda_{\rm flip}$ as
function of the capacitances and Josephson coupling.

%%%%%%%%%%%%%%%%%%%%%%%%%%%%%%%%%%%%%%%%%%%%%% 
\begin{figure*}[!th]
\centering
%\advance\leftskip-0.25cm
\includegraphics[width=1\textwidth]{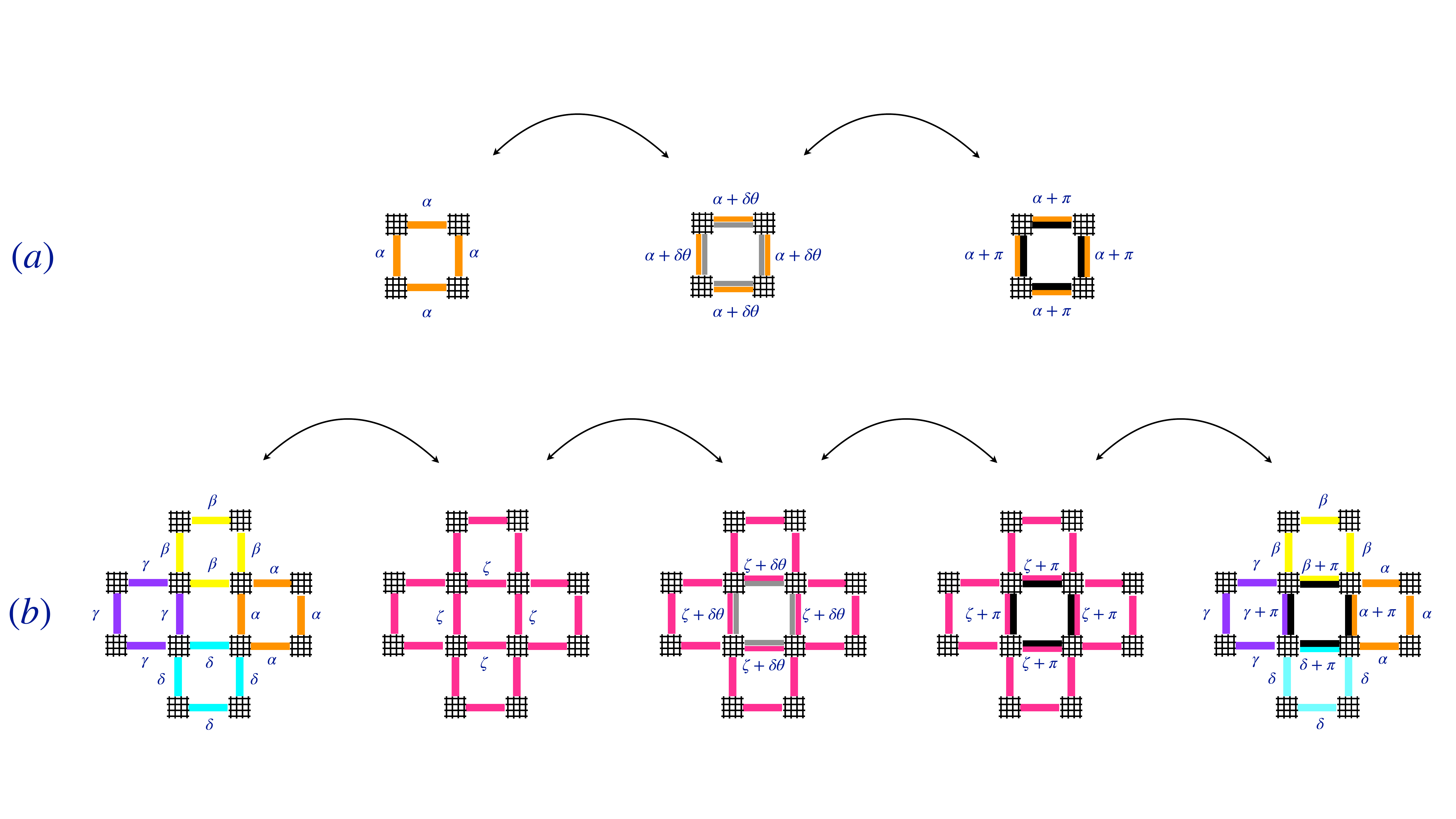}
\caption{Two types of elementary plaquettes with different plaquette
  flipping matrix elements. These two types appear as a consequence of
  the crystal of small loops shown in Fig.~\ref{fig:entropy}. In (a)
  the plaquette contains the elementary $U(1)$ phase variable loop, in
  which all four phases on the edges are equal to some common angle
  $\alpha$. In (b) the plaquette is surrounded by four elementary
  $U(1)$ phase loops, with four different angles
  $\alpha, \beta, \gamma$ and $\delta$ along the edges. In both cases
  adding $\pi$ to each leg around the plaquette costs no Josephson
  energy. The intermediate steps illustrate the path for which one
  goes from the initial to the final configuration without incurring
  any Josephson energy penalty.}
\label{fig:flip} 
\end{figure*}
%%%%%%%%%%%%%%%%%%%%%%%%%%%%%%%%%%%%%%%%%%%%%%

The effective toric code Hamiltonian consisting of the star term
Eq.~\eqref{eq:star-term} and plaquette term
Eq.~\eqref{eq:plaquette-term} is derived by starting from crystal of
small loops shown in Fig.~\ref{fig:entropy}. For illustration, in
Fig.~\ref{fig:toric} we superpose to Fig.~\ref{fig:entropy} the
location of the effective toric spins (at the gauge wires), and the
star ($Z$) and plaquette ($X$) operators acting on the $\mathbb Z_2$
degrees of freedom. To obtain the plaquette flipping amplitudes, we
look at the superconducting phase fluctuations around two types of
elementary plaquettes: (a) plaquettes which contain an elementary
$U(1)$ phase variable loop, in which all four phases on the edges are
equal to some common angle $\alpha$; and (b) plaquettes that are
surrounded by four elementary $U(1)$ phase loops, with four different
angles $\alpha, \beta, \gamma$ and $\delta$ along the edges. [The
phases on the matter wires follow those of the gauge wires according
to Eq.~\eqref{eq:min1}.] These two cases are illustrated in
Fig.~\ref{fig:flip}, on left. Shifting the phases around the four
edges of these plaquettes by $\pi$ does not alter the Josephson
energy, and these configurations are shown on the right part of
Fig.~\ref{fig:flip}. A situation in between the flipped and not
flipped cases is shown on the middle part of the figure, for an
intermediate shift angle $\delta\theta$. The intermediate
configuration for case (a) does not incur an additional Josephson
energy cost for any angle $\delta\theta$, because one can vary this
angle and always remain in the minimum energy configurations
illustrated in Fig.~\ref{fig:JJ_lattice}(c), on the left. In case (b)
there would be a cost if the angles $\alpha, \beta, \gamma$ and
$\delta$ are held fixed. However, there is always a path that incurs
no Josephson energy cost, illustrated by the intermediate steps in
Fig.~\ref{fig:flip}. This path corresponds to changing the four angles
$\alpha, \beta, \gamma$ and $\delta$ on the neighboring plaquettes to
a common value $\zeta$, then changing the shift angle $\delta\theta$
from 0 to $\pi$, and finally returning from $\zeta$ to the original
angles $\alpha, \beta, \gamma$ and $\delta$. Thus, in both cases case
(a) or (b) there is no intermediate Josephson energy cost (i.e., no
classical energy barrier).

However, the absence of a classical Josephson barrier does not mean
that flipping the plaquette is unopposed; quantum fluctuations give
rise to an effective barrier. Notice that, in traversing the path in
Fig.~\ref{fig:flip}(b), one goes from a 4-dimensional space (defined
by the phases $\alpha, \beta, \gamma$ and $\delta$) to another
4-dimensional space where the links are shifted in the middle
plaquette by $\pi$. These two 4-dimensional regions are connected by a
2-dimensional constriction (defined by $\zeta$ and
$\delta\theta$). This constriction of dimensionality leads to level
quantization. The resulting $\delta\theta$-dependent confinement
produces an effective barrier along the $\delta\theta$ direction. The
height of this barrier can be estimated by treating the transverse
motion to $\delta\theta$ as a harmonic oscillator whose potential
energy is of order $J$ and whose kinetic energy is controlled by an
effective capacitance $C$ which is a function of the physical
capacitances of the system. The energy spacing for this
harmonic oscillator is the characteristic frequency
$\omega=\sqrt{J/C}$. Notice that this energy vanishes in the limit
$C\to \infty$, so the effective barrier goes to zero in the classical
limit of infinite capacitances, in agreement with our argument that
the transitions in Fig.~\ref{fig:flip} cost no Josephson energy.

A standard WKB approximation using the effective barrier $\sqrt{J/C}$
with kinetic energy at scale $1/C$ leads to the scaling form~\cite{Garry}
\begin{align}
  \lambda_{\rm flip}
  \sim 
  J\;(JC)^{-k}\;e^{-K\;(JC)^{1/4}}
  \;.
  \label{eq:lambda_flip-b}
\end{align}
The precise size of the gap depends on the numerical constants $k,K$
and prefactors in Eq.~\eqref{eq:lambda_flip-b}. Nevertheless, notice
that the exponent depends on the {\it quartic} root of $JC$, a more
favorable scaling than the usual square root behavior encountered in
other proposals to realize topological phases using superconducting
quantum circuits~\cite{Ioffe1, Ioffe2, Doucot2005, Ioffe2003,
  Gladchenko, Doucot2012}. This qualitative difference is a result of
the absence of a classical Josephson energy barrier in our system,
which is itself a consequence of the combinatorial gauge
symmetry. Moreover, because the combinatorial gauge symmetry is exact
for all values of the coupling $J$ and the capacitances, the existence
of a topological phase is not limited only to the $JC\gg 1$ regime
where the WKB approximation holds, as is the case in previous
proposals where the corresponding symmetry is purely
emergent~\cite{Ioffe1, Ioffe2, Doucot2005, Ioffe2003, Gladchenko,
  Doucot2012}. This opens the possibility of achieving much larger
gaps by reducing $JC$, as long as the system does not transition to
another phase.

Detailed circuit models are needed to identify the effective couplings
and the shape of the potentials discussed above. Preliminary
calculations~\cite{Jamie} for the full $4\times 4$ lattice of waffles
shown in Fig.~\ref{fig:entropy} that include $16 \times 4 = 64$ matter
wires, $40$ gauge wires, their self-capacitances, cross-capacitances,
and the Josephson junction barrier capacitances $C_{\rm J}$ show that
the effective capacitance $C$ above is controlled to leading order by
$C_{\rm J}$, and that the frequency $\omega$ above is given to leading
order by the Josephson plasma frequency $\sqrt{J/C_J}$. Fully
quantitative calculation of $\lambda_{\rm flip}$ (or the gap), the
limits on its size, and its robustness to disorder and noise are
important next steps which we leave for future work.

%In Appendix~\ref{sec:appendix-coordinates} we present a
%circuit-element formulation of the problem that could serve as the
%starting point for a quantitative modeling that addresses the question
%of the maximum $\lambda_{\rm flip}$ size in future work.

In summary, here we showed that finite capacitances lead to a quantum
$\mathbb Z_2$ loop model
\begin{align}
  H^\text{eff} =
  H^\text{eff}_{\rm star}
  +
  H^\text{eff}_{\rm plaquette}
  \;,
  \label{eq:toric}
\end{align}
where $H^\text{eff}_{\rm star}$ and $H^\text{eff}_{\rm plaquette}$ are
given by Eqs.~\eqref{eq:star-term} and~\eqref{eq:plaquette-term},
respectively. In other words, we generated the toric or surface code
Hamiltonian in the superconducting array. Therefore the
superconducting circuit we introduced can serve as a platform for
building topological qubits.

We close this section by commenting that there is a possibility that
the topological phase may even survive the limit of large charging
energies if voltage biases are tuned so two nearly degenerate charge
states are favored in {\it both} matter and gauge wires. In this limit
we reach an interesting spin-1/2 system with two-body interactions and
an exact ${\mathbb Z}_2$ gauge symmetry. We describe this ``WXY''
model in Appendix~\ref{sec:WXY}, and discuss open questions associated
with it.

%%%%%%%%%%%%%%%%%%%%%%%%%%%%%%%%%%%%%%%%%%%%%%

%%%%%%%%%%%%%%%%%%%%%%%%%%%%%%%%%%%%%%%%%%%%%%
\section{Superconducting circuit realization}\label{sec:realization}

We now discuss how the system shown in Figs.~\ref{fig:JJ_lattice}(a)
and (b) can be realized in practice. In addition to implementing the
Josephson potential described by Eqs.~\eqref{eq:JJ-J} and
~\eqref{eq:W}, our circuit must also maintain the required symmetry of
the Hamiltonian in the presence of unavoidable experimental disorder
in circuit parameters. This disorder results both from static
imperfections in physical parameters such as Josephson junction sizes
and capacitances (both discussed below), as well as the presence of
nonstationary ($\sim$1/f) microscopic noise in flux and charge that
occur ubiquitously in superconducting circuits
\cite{chargenoisemooij,chargenoisekuzmin,chargenoisemartinis,ithier,nakamura,bylander,MITLLflux}. Of
course, in the presence of such disorder, no real-world circuit can
ever exhibit perfect combinatorial symmetry, and the success of our
proposals will rely on keeping the residual disorder that cannot be
removed by design, calibration, or adjustment small enough so as to be
only a weak perturbation to the observable physical phenomena of
interest. Also, we stress that the topological phases that we seek to
realize are protected by an energy gap, so the residual disorder only
needs to be suppressed but not necessarily eliminated entirely; as
long as the residual imperfections can be treated perturbatively, they
do not destroy the topological state.

%%%%%%%%%%%%%%%%%%%%%%%%%%%%%%%%%%%%%%%%%%%%%%

\begin{figure*}[!th]
\centering
\includegraphics[width=0.8\textwidth]{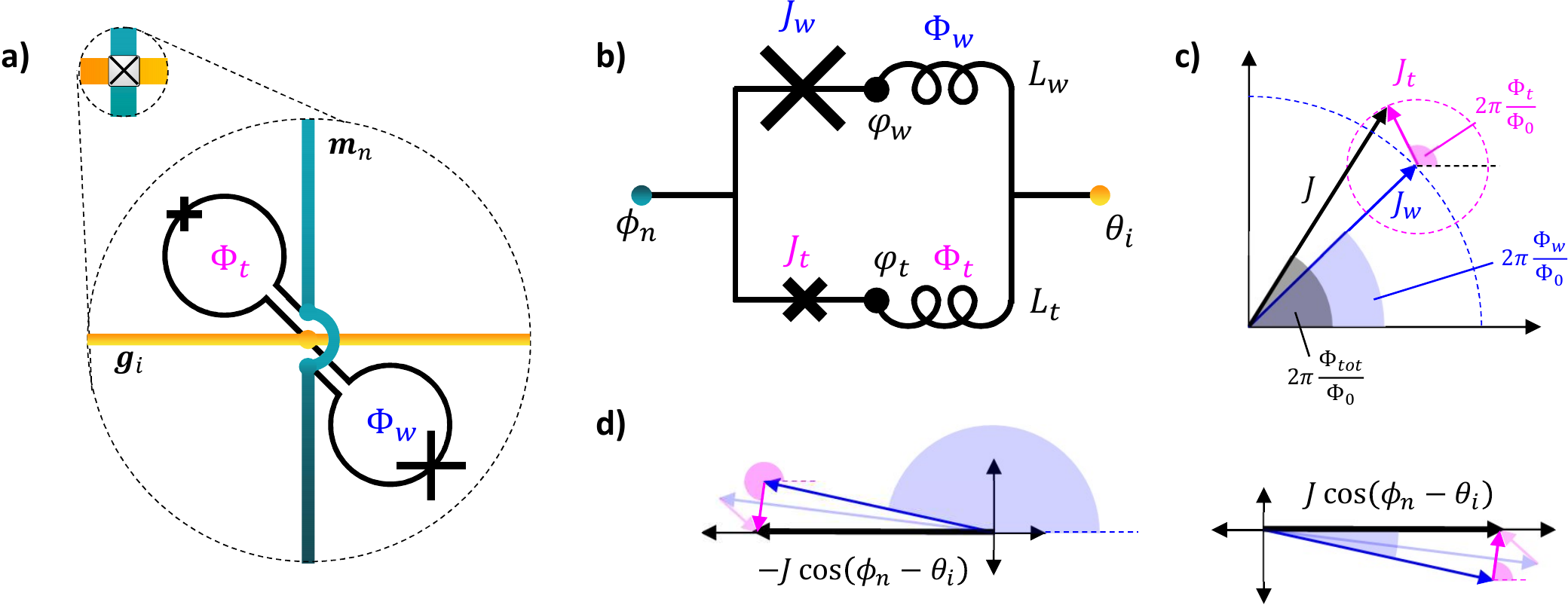}
\caption{Realistic circuit for a single site of the lattice in Fig.~\ref{fig:JJ_lattice}. Panel (a) shows how each Josephson junction in Fig.~\ref{fig:JJ_lattice}(a) is replaced by an asymmetric DC SQUID (b), whose two arms are individually flux-biased with fluxes $\Phi_w$ and $\Phi_t$. Panel (c) illustrates how the effective total Josephson potential for the SQUID can be viewed schematically with a phasor diagram, in which the two Josephson energies $J_{\rm w}$ and $J_{\rm t}$ can be viewed as adding vectorially. Panel (d) then shows how tuning the two fluxes can be used to realize the correct $\pm1$ elements of the $W$ matrix in Eq.~\ref{eq:W}. Shaded arrows in these panels indicate how adjustments to these fluxes can also be used to null out spurious, small variations in the Josephson energies due to fabrication imperfections. }
\label{fig:waffle_circuit} 
\end{figure*}

%%%%%%%%%%%%%%%%%%%%%%%%%%%%%%%%%%%%%%%%%%%%%%

\subsection{Josephson potential}
\label{sec:JJpotential}

The first and most obvious task in formulating an
experimentally-realistic circuit is to produce the Josephson potential
of Eq.~\eqref{eq:JJ-J} with the $W$ given in Eq.~\eqref{eq:W}. To do
this we can exploit the fact that a $c$-number offset of $\pi$ of the
gauge invariant phase difference across a Josephson junction
effectively reverses the sign of its Josephson energy:
$J\cos{(\phi+\pi)}=-J\cos{\phi}$. Such offsets can be easily realized
in superconducting circuits with closed loops using external magnetic
flux, due to the Meissner effect. Although the ``waffle" geometry
naturally presents us with such closed loops in the form of the
plaquettes, each interrupted by four junctions, it is readily seen
that applying flux through these loops will not allow us to achieve to
desired outcome: for each plaquette containing nine loops we must
independently control sixteen $c$-number phase offsets.  (Note that in
the presence of flux noise we cannot hope to take advantage of any
clever geometric scheme exploiting the fact that many of the offsets
are the same; we must require that each $c$-number offset can be
independently controlled and can be used to null out spurious
quasi-static noise.) In addition, relationship between fluxes
threading the plaquettes and parameters in the circuit Hamiltonian
will be complex and nonlinear, not only because wire segments are
shared by numbers of loops, but also higher-order effects such as
spatially non-uniform Meissner screening of the external fields and
imperfect symmetry of individual wire segments' self-inductances.

A viable circuit scheme for achieving the required Hamiltonian control
is shown in Fig.~\ref{fig:waffle_circuit}. First, instead of threading
flux through the plaquette loops, one can use ancillary loops that
replace the single Josephson junction connecting the two wires at each
crossing, as depicted in Fig.~\ref{fig:waffle_circuit}(a). Each of
these loops contains two independently-biased (via fluxes $\Phi_{\rm w}$ and
$\Phi_{\rm t}$) ``arms,'' each of which contains one Josephson junction,
with the two junctions differing in size by a large factor (chosen, as
we discuss below, based on the width of the $J$ distribution for
nominally identical junctions due to fabrication process
variation). The resulting connection between every pair of crossing
wires is then a highly-asymmetric DC SQUID (direct-current
superconducting quantum interference device), as shown in
Fig.~\ref{fig:waffle_circuit}(b), which can be used to control the
tunneling of Cooper pairs between the two wires. We note that, in this
circuit, it will still be experimentally necessary to control the
fluxes through the plaquettes; however, this control will consist
purely of ``magnetic shielding," in that we want all plaquette fluxes
to be zero. Fig.~\ref{fig:waffle_circuit}(c) illustrates how the two
control parameters $\Phi_{\rm w}$ and $\Phi_{\rm t}$ are used, graphically
representing the two Cooper pair tunneling amplitudes as phasors,
whose magnitudes are given by the two Josephson energies, and whose
angles in the complex plane are given by the two external fluxes
$\Phi_{\rm w}$ and $\Phi_{\rm t}$. In this simplified picture, the total Josephson
potential can be approximated (neglecting the finite geometric
inductance of the two arms) as:
\begin{align}
&
  H_{J,ni} = -J\cos{\left(\phi_n-\theta_i+2\pi\frac{\Phi_{\rm tot}}{\Phi_0}\right)}
  \;,
  \label{eq:Jtotmain}
\end{align}
with the definitions:

\begin{align}
  J &\equiv |J_{\rm w}e^{2\pi i\Phi_{\rm w}/\Phi_0}+J_{\rm t}e^{2\pi i\Phi_{\rm t}/\Phi_0}|\nonumber\\
  2\pi\frac{\Phi_{\rm tot}}{\Phi_0} &\equiv {\rm Arg}[J_{\rm w}e^{2\pi i\Phi_{\rm w}/\Phi_0}+J_{\rm t}e^{2\pi i\Phi_{\rm t}/\Phi_0}]
  \;,
  \label{eq:Jtotparams}
\end{align}
where the effective Josephson energy is given by the norm of the
vector sum of the two phasors, and the $c$-number offset to its
gauge-invariant phase difference by the argument of that vector sum (see appendix~\ref{sec:squid} for details). 
    
The solid arrows in panel~\ref{fig:waffle_circuit}(d) then show how for appropriate
choices of the fluxes the potential can be set with phase offsets of 0
(right) and $\pi$ (left). Finally, the lightly-shaded arrows in
panel~\ref{fig:waffle_circuit}(d) indicate how a desired amplitude
$\pm J$ can be obtained and made uniform across different junctions
even in the presence of static variations in Josephson energy (due to
fabrication process variation of junction size or critical current
density). By choosing the smaller junction size based on the maximum
amplitude of these variations (which for a state-of-the art
shadow-evaporated Aluminum Josephson junction process can be as low as
a few percent~\cite{JJsize}), we can ensure that the circuit is tunable enough to
null them out. We remark that this could be a nontrivial process
experimentally, and may require additional ancillary observables to be
integrated into the circuit to make this calibration feasible,
depending on the quantitative level of symmetry required for a given
experimental goal.

In closing this section, we note that one could also in principle use
``$\pi$-junctions," Josephson junctions with a ferromagnetic
barrier,~\cite{Frolov2004,yamashita2005,ustinov2010} to achieve the
required phase offsets. However, this would not allow the phase shifts
to be controlled {\it in situ} to minimize breaking of the
combinatorial symmetry by fabrication variations, as we have just
described, and is a less well-developed technology than junctions with
a conventional dielectric barrier.

%%%%%%%%%%%%%%%%%%%%%%%%%%%%%%%%%%%%%%%%%%%%%%
\subsection{Electrostatic potential}

\begin{figure}[!th]
\centering
\includegraphics[width=0.4\textwidth]{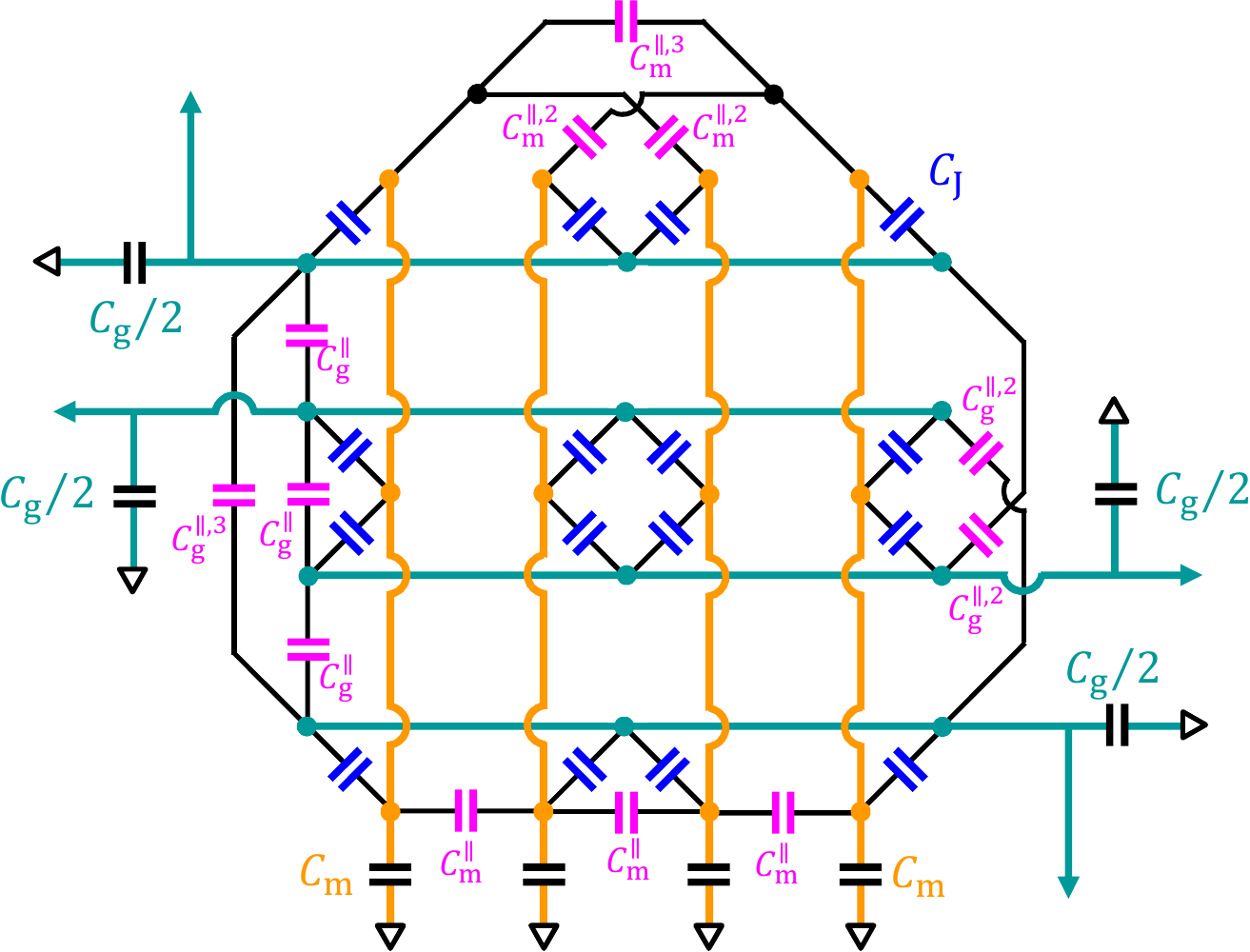}
\caption{Illustration of the capacitances present in a single site of the JJ array. Green thick lines are gauge wires, and orange thick lines are matter wires. }
\label{fig:Cmat} 
\end{figure}

Figure~\ref{fig:Cmat} illustrates the relevant capacitances for a
single site containing 4 gauge (green lines) and 4 matter (orange
lines) wires. By far the largest in magnitude among these are the
Josephson junction barrier capacitances $C_{\rm J}$ (shown in blue),
scaling with the junction area like the Josephson energy $J$. Typical
magnitudes of these for shadow-evaporated Aluminum junctions are
$\sim40-80\;f$F$/\mu m^2$, with the corresponding $J$ values ranging
from $\sim k_{\rm B}\times10-200$ K$/\mu m^2$. Each wire also has a
self-capacitance to ground, shown in black, where we have defined the
gauge wire self-capacitances as $C_{\rm g}/2$ since each of these
wires spans two sites. Finally, there are parasitic capacitances
between parallel wires, shown with magenta in the figure. For adjacent
wires this quantity is labeled $C^{||}$, while the smaller parasitics
between next nearest and between the outside pair of wires are
labeled $C^{||,2}$ and $C^{||,3}$, respectively. We can safely ignore
the parasitics between matter and gauge wires, since these always
appear directly in parallel with the much larger $C_{\rm J}$. (See
Appendix ~\ref{sec:Cmat} for the capacitance matrix.)

%%%%%%%%%%%%%%%%%%%%%%%%%%%%%%%%%%%%%%%%%%%%%% 
%%%%%%%%%%%%%%%%%%%%%%%%%%%%%%%%%%%%%%%%%%%%%%

In section~\ref{sec:JJpotential} we discussed the requirements to
realize the magnetic potential of Eqs.~(\ref{eq:JJ-J})
and~(\ref{eq:W}) which exhibits combinatorial gauge symmetry. The
question also arises, however, whether this symmetry can be broken in
any important ways by the \textit{electrostatic} part of the
Hamiltonian, Eq.~(\ref{eq:HK}). This corresponds to non-invariance of
$\mathbf{C}^{-1}$ under permutations of the matter wires within each
site. Referring to Fig.~\ref{fig:Cmat}: while it is reasonable to
assume that the self-capacitance of the matter wires $C_{\rm m}$ can
be made symmetric within each site, the parasitic capacitances between
matter wires $C_{\rm m}^{||}$, $C_{\rm m}^{||,2}$, and
$C_{\rm m}^{||,3}$ will not naturally be equal, and therefore
will break the symmetry.

\begin{figure}[!tbh]
\centering
\includegraphics[width=0.25\textwidth]{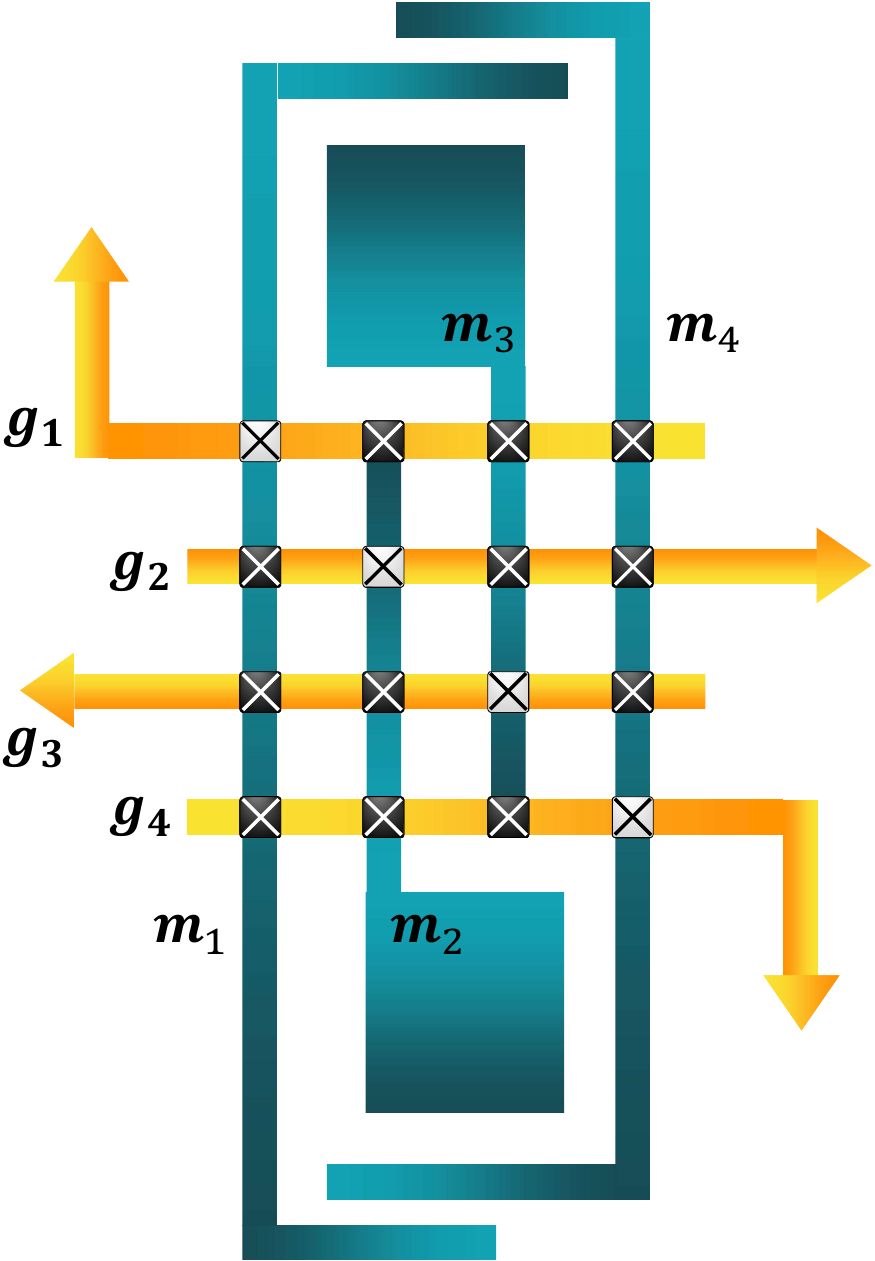}
\caption{Schematic for restoring combinatorial symmetry to the inverse capacitance matrix of a site.}
\label{fig:matter_wires} 
\end{figure}

However, these parasitic capacitances do not appear in the tunneling
energy to leading order~\cite{Jamie}, so that from the perspective of
low-energy, static emulation of the toric code, we are justified in
neglecting this small symmetry breaking. That said, it is still
possible that these effects could become important when the system's
dynamic response to noise is considered, in the context of topological
protection of quantum information. Should this turn out to be the
case, we show schematically in Figure~\ref{fig:matter_wires} that with
appropriate electrostatic design, the parasitic capacitances could be
symmetrized so as to null out the combinatorial symmetry-breaking.

%%%%%%%%%%%%%%%%%%%%%%%%%%%%%%%%%%%%%%%%%%%%%%
%%%%%%%%%%%%%%%%%%%%%%%%%%%%%%%%%%%%%%%%%%%%%%

Another effect in the electrostatic Hamiltonian which could break the
combinatorial symmetry would be a spurious asymmetry in the
capacitances across the DC SQUID coupling elements between matter and
gauge wires (arising from fabrication process variation of junction
sizes). Although such imperfections contribute in leading order to the
diagonal elements of $\mathbf{C}^{-1}$, the resulting breaking of the
symmetry can be expected to be quite small, for two reasons. First,
unlike the fabrication variations in the Josephson energy, which
depend exponentially on the dielectric barrier thickness, junction
capacitance $C_{\rm J}$ depends only inversely on this thickness, so
that the resulting variations are even smaller than that observed in
$J$. Second, the numerical coefficient of the linear correction term
breaking the symmetry is small (1/8 in the simplest circuit
model~\cite{Jamie}), pushing the resulting
expected fractional variation between diagonal charging energy terms
to the $\sim10^{-3}$ level for junction size uniformity of a few
percent\cite{JJsize}.

%%%%%%%%%%%%%%%%%%%%%%%%%%%%%%%%%%%%%%%%%%%%%%

\section*{Summary}
We have proposed a superconducting quantum circuit based on Josephson
junction arrays that realizes combinatorial gauge symmetry. This
symmetry is both local and exact and leads to interesting loop phases
with topological order. We have argued that the model admits a gapped
quantum topological phase which should be stable for a wide range of
parameters. The general framework laid out here offers a promising
path to engineering exotic many-body states in the laboratory and to
realizing a platform for topological quantum computation.

\section*{Acknowledgments}
We thank Sergey Frolov, Garry Goldstein, Andrei Ruckenstein, Zhi-Cheng
Yang, and Hongji Yu for useful discussions and constructive
criticism. This work is supported in part by the NSF Grant
DMR-1906325. C.~C. thanks the hospitality of the NSF Quantum Foundry
at UCSB during the initial stages of this work.  A.K. was funded by
the Assistant Secretary of Defense for Research, Engineering under Air
Force Contract No. FA8721-05-C-0002. The views and conclusions
contained herein are those of the authors and should not be
interpreted as necessarily representing the official policies or
endorsements, either expressed or implied, of the US Government.

\appendix

\section{Plaquette operators} 

The local combinatorial gauge symmetry allows us to construct local
conserved quantities -- plaquette operators on each plaquette $p$:
\begin{align}
  G_p = \prod_{s \in p} {\mathcal{L}}^{(\hA)}_s\prod_{i\in p}
  {\mathcal{R}}^{(\hB)}_i~.
\label{eq:plaquette}
\end{align}
The product of the ${\mathcal{R}}^{(\hB)}_i$ around the plaquette
flips both legs at each corner site $s$. An example is shown in the
inset of Fig.~\ref{fig:JJ_lattice}(c). $\mathcal{L}^{(\hA)}_s$ is the
companion operator that permutes and flips the matter fields at each
corner site per Eqs.~(\ref{eq:LR}) and ~(\ref{eq:U}). Any two $L$
matrices commute and therefore the plaquette operators do as well,
$[G_p,G_{p^\prime}]=0$. Finally, gauge symmetry guarantees that $G_p$
is conserved for all $p$: $[H, G_p]=0$.

\section{Loop fugacity of continuous variables}
\label{sec:appendix_fugacity}

Here we clarify the argument in the main text regarding the zero
temperature partition function of the $U(1)$ loop component. Recall
that our phases are defined mod $\pi$ and not mod $2\pi$, but the
difference is simply a factor of 1/2 in the fugacity, so below we
present the argument for the mod $2\pi$ case and return to this issue
at the end of this appendix section.

Consider a loop $\mathcal{C}$ with length $p$, $p$ continuous
variables $\theta_i, i=1,\dots p$, and a Boltzmann factor that
penalizes configurations where consecutive variables are different
($\theta_p$ is consecutive to $\theta_1$).  We can write the
contribution of this loop to the partition function as
\begin{widetext}
\begin{align}
  Z_\mathcal{C} =
  \int_0^{2\pi} \frac{d\theta_1}{2\pi}\,
  \frac{d\theta_2}{2\pi}\dots
  \dots
  \frac{d\theta_p}{2\pi}
  \;
  e^{-K[1-\cos(\theta_1-\theta_2)]}\;
  e^{-K[1-\cos(\theta_2-\theta_3)]}\;
  \dots
  e^{-K[1-\cos(\theta_{p-1}-\theta_{p})]}\;
  e^{-K[1-\cos(\theta_{p}-\theta_{1})]}\;
  \;,
\end{align}
where the Boltzmann factor that imposes the constraint is $e^{-K[1-\cos(\theta_i-\theta_{i+1})]}$. The overall ground state energy has been left out because it is identical for all configurations on the lattice. As $K\to\infty$ we can replace the factors by Gaussian approximations:
\begin{align}
  \label{eq:Z_loop_appendix}
  Z_{\mathcal C}
  &\approx
  \int_0^{2\pi} \frac{d\theta_1}{2\pi}\,
  \frac{d\theta_2}{2\pi}\dots
  \dots
  \frac{d\theta_p}{2\pi}
  \;
  e^{-\frac{K}{2}(\theta_1-\theta_2)^2}\;
  e^{-\frac{K}{2}(\theta_2-\theta_3)^2}\;
  \dots
  e^{-\frac{K}{2}(\theta_{p-1}-\theta_{p})^2}\;
  e^{-\frac{K}{2}(\theta_{p}-\theta_{1})^2}\;
  \nonumber\\
  &\approx
  \int_0^{2\pi} \frac{d\theta_1}{2\pi}\,
  \frac{d\theta_2}{2\pi}\dots
  \dots
  \frac{d\theta_p}{2\pi}
  \;
  [\sqrt{\frac{2\pi}{K}}\;\delta(\theta_1-\theta_2)]\;
  [\sqrt{\frac{2\pi}{K}}\;\delta(\theta_2-\theta_3)]\;
  \dots
  [\sqrt{\frac{2\pi}{K}}\;\delta(\theta_{p-1}-\theta_{p})]\;
  e^{-\frac{K}{2}(\theta_{p}-\theta_{1})^2}\;
  \nonumber\\
  & =
  \left(\sqrt{\frac{1}{2\pi K}}\right)^{p-1}
  \int_0^{2\pi} \frac{d\theta_1}{2\pi}
  \;,
\end{align}
\end{widetext}
where we replaced the Gaussian approximations by the delta functions
for all links with the exception of the last because it is
automatically enforced by the other $p-1$ constraints. The one less
power of the factor $\sqrt{1/2\pi K}$ has its
origin in the last link, which closes the loop.

In a fully packed lattice model, each term in the partition function
is an integral over $n_B$ bonds, where $n_B$ is the total number of
bonds on the lattice. Therefore each loop configuration will
contribute the factor $\left(\sqrt{1/2\pi K}\right)^{n_B-n_\ell}$, where
$n_\ell$ is the number of loops in the configuration.  Ignoring the
overall factor for number of bonds, we can identify the loop fugacity
$\lambda$ as
\begin{align}
  \lambda = \sqrt{2\pi K}\to\infty\quad {\rm if}\;K\to\infty
    \;.
\end{align}

We can think of this result formally as the integration over one
redundant delta function since only $p-1$ delta functions are required
to enforce a constraint around a loop with perimeter $p$; the
remaining delta function is evaluated at 0, giving the value
``$\delta(0)$'' to the fugacity.

More intuitively, the result follows the simple expectation that we
have a continuous phases (infinitely many colors) associated with each
loop.

We now return to the issue that our phases are defined mod $\pi$ and
not mod $2\pi$. Changing all the $\int_0^{2\pi}\frac{d\theta_i}{2\pi}
\to \int_0^{\pi}\frac{d\theta_i}{\pi}$ in
Eq.~(\ref{eq:Z_loop_appendix}) changes the result for the fugacity by
a factor of 1/2, i.e., we replace the $\lambda$ found above by
$\lambda/2$. Of course, none of the discussion above changes as
$\lambda\to\infty$. Nonetheless, this factor reinforces the simple
intuitive interpretation of the continuous angles representing
infinitely many colors: half the continuous angles correspond to half
the infinitely many colors, as expressed by the scaling of $\lambda$
by 1/2.

%%%%%%%%%%%%%%%%%%%%%%%%%%%%%%%%%%%%%%%%%%%%%%

%%%%%%%%%%%%%%%%%%%%%%%%%%%%%%%%%%%%%%%%%%%%%%
\section{``WXY Model": limit of small capacitances in all wires}
\label{sec:WXY}

We present another limit of the superconducting wire array that is
equivalent to a spin-1/2 system with two-body interactions and an
exact ${\mathbb Z}_2$ gauge symmetry. Consider the limit where both
the matter and gauge capacitances are small, and voltage biases are
tuned so two nearly degenerate charge states are favored in each
matter and gauge wire. In this limit, we reach an interesting spin
model that could potentially have a gap of order $E_J$, as we argue
below.

In the limit when the wires become two-level systems, we can deploy
spin-1/2 raising and lowering operators via the replacements
$e^{\pm\i\, \phi_n}\to \mu^{\pm}_n$ and
$e^{\pm\i\,\theta_i}\to \sigma^{\pm}_i$. In terms of combinatorial
gauge symmetry we can identify the small capacitance version of
Eq.(\ref{eq:JJ_full}):
\begin{align}
  \hA^{}_n = \mm_n 
  \quad
  &\text{and}
  \quad
  \hB^{}_i=\sm_i
\nonumber\\
  U^{(L)}_n = e^{\i\frac{\pi}{2}\, \mz_n} = \i\mz_n
  \quad
  &\text{and}
  \quad
  U^{(R)}_i = e^{\i\frac{\pi}{2}\, \sz_i} = \i\sz_i
  \;.
  \label{eq:U_quantum}
\end{align}
The Hamiltonian takes the form reminiscent of the standard quantum
XY-model, but again with the crucial Hadamard symmetry:
\begin{align}
  H_J = -J\;\sum_s \sum_{n,i\in s} W_{ni}\;
  \left(
  \mx_n\;\sx_i
  +
  \my_n\;\sy_i
  \right)
  \;.
  \label{eq:WXY}
\end{align}
We refer to this model as ``WXY". If the wires are biased slightly
away from the degenerate point, kinetic terms of the form $\mz_n$ and
$\sz_i$ appear. These terms commute with both $U^{(L)}_n$ and
$U^{(R)}_i$, and if the couplings are uniform they satisfy the
permutation part of the combinatorial gauge symmetry. However, these
kinetic terms are not required as quantum dynamics is present at the
outset in the WXY-model of Eq.~(\ref{eq:WXY}).

Is its low-energy spectrum gapped? Answering this question is outside
the scope of this work, and may require a detailed numerical
study. [We note that the quantum WXY-model of Eq.~(\ref{eq:WXY}) has a
sign problem, so numerical studies would require methods such as the
Density Matrix Renormalization Group (DMRG).] If the model does turn
out to be gapped, the only energy scale in the Hamiltonian is $J$, and
therefore such gap would be rather large. Given that the model has an
exact ${\mathbb Z}_2$ symmetry, we believe it is an interesting spin
model to study even if it is gapless.

%%%%%%%%%%%%%%%%%%%%%%%%%%%%%%%%%%%%%%%%%%%%%%

\section{Capacitance matrix}\label{sec:Cmat}

\begin{widetext}
\begin{align}
  \mathbf{C}_{\rm s} =
  \begin{psmallmatrix}
    C_{\rm m}+4C_{\rm J}+C_{\rm m}^{||}	&	-C_{\rm m}^{||}	&	-C_{\rm m}^{||,2}	&	-C_{\rm m}^{||,3} & -C_{\rm J} & -C_{\rm J} & -C_{\rm J} & -C_{\rm J}\\
    +C_{\rm m}^{||,2}+C_{\rm m}^{||,3}&\\
    \\
    -C_{\rm m}^{||}	&	C_{\rm m}+4C_{\rm J}	&	-C_{\rm m}^{||}	&	-C_{\rm m}^{||,2} & -C_{\rm J} & -C_{\rm J} & -C_{\rm J} & -C_{\rm J}\\
    &+2C_{\rm m}^{||}+C_{\rm m}^{||,2}\\
    \\
    -C_{\rm m}^{||,2}	&	-C_{\rm m}^{||}	&	C_{\rm m}+4C_{\rm J}	&	-C_{\rm m}^{||} & -C_{\rm J} & -C_{\rm J} & -C_{\rm J} & -C_{\rm J}\\
    &&+2C_{\rm m}^{||}+C_{\rm m}^{||,2}\\
    \\
    -C_{\rm m}^{||,3}	&	-C_{\rm m}^{||,2}	&	-C_{\rm m}^{||}	&	C_{\rm m}+4C_{\rm J}+C_{\rm m}^{||}& -C_{\rm J} & -C_{\rm J} & -C_{\rm J} & -C_{\rm J}\\
     &&&   +C_{\rm m}^{||,2}+C_{\rm m}^{||,3}&\\
     \\
    -C_{\rm J} & -C_{\rm J} & -C_{\rm J} & -C_{\rm J}& C_{\rm g}/2+4C_{\rm J}+C_{\rm g}^{||}	&	-C_{\rm g}^{||}	&	-C_{\rm g}^{||,2}	&	-C_{\rm g}^{||,3} \\
    &&&&+C_{\rm g}^{||,2}+C_{\rm g}^{||,3}&\\
    \\
    -C_{\rm J} & -C_{\rm J} & -C_{\rm J} & -C_{\rm J}& -C_{\rm g}^{||}	&	C_{\rm g}/2+4C_{\rm J}	&	-C_{\rm g}^{||}	&	-C_{\rm g}^{||,2} \\
    &&&&&+2C_{\rm g}^{||}+C_{\rm g}^{||,2}\\
    \\
    -C_{\rm J} & -C_{\rm J} & -C_{\rm J} & -C_{\rm J}&-C_{\rm g}^{||,2}	&	-C_{\rm g}^{||}	&	C_{\rm g}/2+4C_{\rm J}	&	-C_{\rm g}^{||} \\
    &&&&&&+2C_{\rm g}^{||}+C_{\rm g}^{||,2}\\
    \\
    -C_{\rm J} & -C_{\rm J} & -C_{\rm J} & -C_{\rm J}&-C_{\rm g}^{||,3}	&	-C_{\rm g}^{||,2}	&	-C_{\rm g}^{||}	&	C_{\rm g}/2+4C_{\rm J}+C_{\rm g}^{||}\\
    &&&&&&&   +C_{\rm g}^{||,2}+C_{\rm g}^{||,3}&\\
     \\
  \end{psmallmatrix}
  \label{eq:Cmat}
\end{align}
\end{widetext}

%\section{Combinatorial symmetry and the electrostatic Hamiltonian}\label{sec:Cmatsym}

\section{Magnetic potential of asymmetric DC SQUID coupling elements}\label{sec:squid}

The magnetic potential energy of the asymmetric DC SQUID in Fig.~\ref{fig:waffle_circuit}(b) can be written (taking $L_{\rm w}=L_{\rm t}=L$ and $J_{\rm t}/J_{\rm w}\equiv d_J\ll 1$):

\begin{subequations}
\begin{align}
&
  \frac{U}{J_{\rm w}} =-\cos{\left(\phi_n-\theta_i-\varphi_{\rm ow}+2\pi\frac{\Phi_{\rm w}}{\Phi_0}\right)}\nonumber\\
&  -d_J\cos{\left(\phi_n-\theta_i-\varphi_{\rm ot}+2\pi\frac{\Phi_{\rm t}}{\Phi_0}\right)}\nonumber\\
&+\frac{1}{2e_{LJ}}\left(\varphi_{\rm ow}^2+\varphi_{\rm ot}^2\right)
  \label{eq:Jtot}
\end{align}
where we have defined the displaced oscillator coordinates:
\begin{align}
&\varphi_{\rm ow}\equiv\varphi_{\rm w}-\theta_i-2\pi\frac{\Phi_{\rm w}}{\Phi_0}\nonumber\\
&\varphi_{\rm ot}\equiv\varphi_{\rm t}-\theta_i-2\pi\frac{\Phi_{\rm t}}{\Phi_0}
\end{align}
and the ratio between Josephson and linear inductive energy:
\begin{align}
e_{LJ}\equiv \frac{4\pi^2L}{\Phi_0^2}J_{\rm w}
\end{align}
\end{subequations}

\noindent We can simplify Eq.~\ref{eq:Jtot} by approximating the two (high-frequency) oscillator mode coordinates by the values which minimize the inductive energy with respect to each coordinate. Substituting these values for $\varphi_{ow}$ and $\varphi_{ot}$ back into Eq.~\ref{eq:Jtot}, we obtain, to second order in $d_L$ and $e_{LJ}$:

\begin{align}
&
  \frac{U}{J_{\rm w}}\approx-\left(1-\frac{e_{LJ}^2}{8}\right)\cos{\left(\phi_n-\theta_i+2\pi\frac{\Phi_{\rm w}}{\Phi_0}\right)}\nonumber\\
&  -d_J\cos{\left(\phi_n-\theta_i+2\pi\frac{\Phi_{\rm t}}{\Phi_0}\right)}\nonumber\\
&   +\frac{e_{LJ}}{4}\cos{\left[2\left(\phi_n-\theta_i+2\pi\frac{\Phi_{\rm w}}{\Phi_0}\right)\right]}\nonumber\\
&   -\frac{e_{LJ}^2}{8}\cos{\left[3\left(\phi_n-\theta_i+2\pi\frac{\Phi_{\rm w}}{\Phi_0}\right)\right]}
\end{align}

\noindent The first two terms of this result correspond to the phasor diagram of Fig.~\ref{fig:waffle_circuit}(c), with a renormalization of the effective Josephson energy $J_{\rm w}$ due to the finite loop inductance. 

The last two terms are first and second-order distortions of the
effective Josephson potential by the loop inductance, which can be
viewed as weak two- and three-Cooper-pair tunneling terms. Although
the effect of these terms on the phases of our model are not yet
understood, they can be made small via the parameter $e_{LJ}$. The
extent to which this parameter can be reduced will be determined by
how small the loop inductances and Josephson energies can be made,
while retaining the ability to provide sufficient bias flux and
keeping the Josephson energy scale large enough compared to $k_BT$. To
get a rough estimate of this quantity, if we take the reasonable
values: $J_{\rm w}\sim k_B\times 1$K, and $L_w\sim 10$pH, we obtain:
$e_{LJ}\sim 10^{-4}$.

\bibliography{reference}

\end{document}